\documentclass[a4paper,11pt]{JHEP3}
\usepackage{fancyheadings}
\usepackage{ifthen}
\usepackage{amsmath}
\usepackage{epsfig}
\input feynman
\usepackage[errorshow]{tracefnt}
\newcommand\nn{\nonumber}

\newcommand\beal{\begin{align}}
\newcommand\eeal{\end{align}}
\newcommand\benu{\begin{enumerate}}
\newcommand\eenu{\end{enumerate}}
\newcommand\bit{\begin{itemize}}
\newcommand\eit{\end{itemize}}

\newcommand{\slsh}[1]{\displaystyle{\not} #1}

\newcommand{\ee}{\end{equation}}
\newcommand{\bd}{\begin{displaymath}}
\newcommand{\ed}{\end{displaymath}}

\newcommand\al{\alpha}
\newcommand\be{\beta}

\newcommand\q{\theta}
\newcommand\g{\gamma}
\newcommand\la{\lambda}

\newcommand\bq{\bar{\theta}}

\newcommand\bdel{\bar{\partial}}

\newcommand\del{\partial}
\newcommand\h{\frac{1}{2}}

\newcommand\GV{{\cal G}}

\newcommand\tr{\mathrm{tr}}

\title{$R^4$, purified}

\author{Giuseppe Policastro\\
Arnold Sommerfeld Center for Theoretical Physics\\
Department f\"{u}r Physik, Ludwig-Maximilians-Universit\"at\\
Theresienstr. 37, 80333 M\"unchen, Germany\\
E-mail: \email{policast@theorie.physik.uni-muenchen.de}}

\author{Dimitrios Tsimpis \\
Max-Planck-Institut f\"{u}r Physik --Theorie\\
F\"{o}hringer Ring 6,  80805 M\"{u}nchen, Germany\\
E-mail: \email{tsimpis@mppmu.mpg.de}
}

\abstract{
We derive, using the pure-spinor formalism,  
the complete -- including the fermions -- four-point effective action 
of both type II superstrings to all orders in $\alpha'$, at tree level in 
string loops. 
We find that, in the quartic-field approximation, 
the supergravity Lagrangian can be thought of 
as the tensor product, in a suitable sense,  of two copies of the 
superYang-Mills Lagrangian in ten dimensions. 
The NS-NS three-form enters 
the supergravity 
Lagrangian through a modified connection with torsion. 
As a byproduct, we derive the complete, i.e. to all orders in the $\theta$-expansion, 
closed-string vertex operator in a flat target-space background.
}

\preprint{LMU-ASC 17/06\\ MPP-2006-30}
\begin{document}

\section{Introduction}

The pure-spinor formalism for the superstring introduced a few years ago in \cite{ba} 
(see \cite{bb} for a review), has been a remarkable technical advance, overcoming long-standing 
difficulties with both the Green-Schwarz and the RNS formalisms. 
Pure spinors were originally introduced in ten-dimensional superYang-Mills 
in an attempt to construct an off-shell formulation \cite{sym}. 
It was subsequently noted that the on-shell constraints in both ten-  
and eleven-dimensional 
supersymmetric theories 
\cite{purehoweten, purehoweeleven} 
can be thought of as pure-spinor integrability conditions. 
Eventually Berkovits incorporated these early insights into a full-fledged formulation of string theory.
Berkovits' proposal is self-consistent and unambiguous, and 
has already 
passed a number of nontrivial tests \cite{b, bc, bca, bd, bda, be, bf}. 
However the origin of 
some of the prescriptions of the pure-spinor formalism remains largely mysterious. 
Perhaps 
most significantly, it has proven impossible so far to understand the pure-spinor 
BRST cohomology operator as coming from the gauge-fixing of a 
reparameterization-invariant action
\footnote{In \cite{Aisaka:2005vn} it was shown that the 
pure-spinor formalism can be derived by gauge-fixing 
a reparameterization-invariant action of Green-Schwarz type. So far, however, this procedure 
has not proven useful in understanding  the origin of Berkovits' prescription for 
the scattering amplitudes, since it runs into the well-known problems 
associated with the semi-light-cone gauge.}. 
As a consequence, the computation of scattering amplitudes does not 
follow from a path  integral quantization \`{a} la 
Polyakov, but has to be performed using a set of more-or-less ad-hoc rules. 

There have been several attempts in the literature
to justify these rules and `derive' the formalism 
from different points of view.  An alternative formulation which 
dispenses with using pure spinors by embedding  
Berkovits' theory in a larger theory with additional fields, 
was put forward in \cite{giusa, giusb}. 
This model can then be understood formally as a WZW model with 
${\cal N}=2$ supersymmetry based on a superalgebra 
which is a fermionic central extension of the Poincar\'{e} 
superalgebra of the target space \cite{Grassi:2003kq}. 
One can recover the correct spectrum in a completely covariant 
manner \cite{Grassi:2002tz}. 
Recently it has been shown that the ${\cal N}=2$ algebra can also arise from a different 
set of non-minimal fields \cite{Berkovits:2005bt}.  
For other approaches, see also \cite{Aisaka:2005vn}, \cite{Oda:2005sd}, \cite{Gaona:2005yw}.

%Pure spinors were originally introduced in ten-dimensional superYang-Mills 
%in an attempt to construct an off-shell formulation \cite{sym}. 
%It was subsequently noted that the on-shell 
%constraints in both ten \cite{purehoweten} 
%and eleven dimensions \cite{purehoweeleven} can be thought of 
%as pure-spinor integrability conditions. 

Although it is quite important to 
put the pure-spinor formalism on a sound  conceptual footing, 
there is by now little doubt that it works, and for practical computations 
it is likely to be the most economical one. 
For this reason, in the present paper we adopt a pragmatic approach, and simply 
take advantage of the natural, 
Poincar\'{e}-covariant way in which the fermions and 
the Ramond-Ramond fields are described in the pure-spinor formalism, in order to 
derive the complete -- fermions included -- four-point tree-level effective action 
of both type II superstrings to all orders in $\alpha'$. Our approach makes 
transparent a simple relation, which was already known to hold 
in the NS-NS sector \cite{grosssloan, kehagiasa, kehagiasb}, between the four-point 
gravitational action and the superYang-Mills action in ten dimensions. I.e.  
the former is the tensor product (in a sense which we make 
precise in section \ref{fpam}) of two copies of the latter,
\beal
{\cal L}_{SUGRA}={\cal L}_{SYM}\otimes\widetilde{{\cal L}}_{SYM}~.
\label{tensor}
\end{align}
We stress that this only holds at the level of 
the four-field approximation. 
Equation (\ref{tensor}) is a direct consequence of the tree-level 
relations between open- and closed-string amplitudes, first observed in \cite{klt}. 
We give a general proof of (\ref{tensor}), 
using the Kawai-Lewellen-Tye relations, in section \ref{fpam}. 
We have also verified it explicitly by 
deriving the four-point Lagrangian in both ways: by direct computation of the 
closed-string four-point amplitudes, as well as by `squaring' the SYM Lagrangian.

The consequences of the KLT relations have been investigated by Z. 
Bern and collaborators (see for instance \cite{Bern:2002kj} 
and references therein), who have found a way of considerably 
simplifying perturbative computations in pure gravity by 
systematically exploiting the factorization property and other inputs from string theory. In 
the present paper we focus on the quartic part of 
the full supergravity Lagrangian; we do 
not make any attempt to generalize the constructions of Bern et al 
to the complete type II supergravity -- although that would certainly be interesting. 

In hindsight the present paper could have been written shortly after 
the publication of \cite{klt}: anticipating 
the advance of a covariant formulation of the superstring, one could have 
postulated a manifestly Poincar\'{e}-invariant 
open-string four-point amplitude. This 
amplitude would be the covariantization of the four-point 
RNS or Green-Schwarz amplitude, and would be such that 
it reproduces the known four-field, 
four-derivative correction to the superYang-Mills 
action in ten dimensions. 
Indeed, as we shall see in section \ref{opsamp}, 
the open-string amplitude derived using the pure-spinor 
approach has precisely these properties.  
The closed-string Lagrangian would then follow form 
the tensor product (\ref{tensor}).

As a by-product of the present investigation, 
we are able to derive the complete, i.e. to all orders in the $\theta$-expansion, 
closed-string vertex operator in a flat target-space background. 
The $\theta$-expansion of the 
closed-string vertex operators for type II superstrings in curved backgrounds 
was considered in \cite{gt}, where an iterative algorithm for computing the expansion 
at each order in $\theta$ was presented. The procedure of \cite{gt} is equivalent to 
the normal-coordinate expansion in superspace \cite{mac}, which was first developed   
in the context of  four-dimensional ${\cal N}=1$ supergravity.  
The normal coordinate expansion 
has recently been applied by one of the present authors 
to the case of eleven-dimensional supergravity \cite{curved}.  
As  was observed in that reference, in the linear approximation the 
$\theta$-expansion can be explicitly computed to all orders. This fact was then used 
in \cite{curved} to derive the membrane 
vertex operator in flat target space, to all orders in $\theta$. As we 
show in section \ref{norcor}, an all-order result is also possible
 in the case of the superstring.

There already exists a large amount of literature on higher-order 
$\alpha'$-corrections to ordinary supergravities in ten dimensions, 
see \cite{pvwreview} for a review and a more extensive list of references. In type II 
theories most of our information about these corrections comes 
from perturbative string-theory computations. However, 
at the level of the four-field approximation, 
the eight-derivative correction to the type IIB supergravity 
is implicitly known in superspace: it is 
captured by the chiral integral 
(i.e. the integral 
over half the superspace) of the fourth power of the 
linearized scalar superfield of \cite{hw}. Unfortunately, the complete component 
form of the action 
is somewhat tedious to extract, although it is 
known to reproduce the $t_8t_8R^4$ part of the Lagrangian \cite{nilsson}. 
An interesting diagrammatic technique to perform the superspace integral 
has recently been proposed in \cite{gps}. It seems unlikely to us, however, 
that the computation in the present paper would be made any easier by adopting the methods of 
\cite{gps}. We comment further on the linearized superfield in section \ref{lit}. 

The RR sector of type IIB superstrings was considered by Peeters and Westerberg 
in \cite{pw}. These authors were able to perform the formidable task of obtaining 
the Poincar\'{e}-covariant form of the 
two-graviton, two-RR field scattering amplitudes, working at first order 
in string loops within the RNS formalism. The methods of 
\cite{pw} in handling the RR sector, which make use of earlier techniques 
developed some time ago by Atick and Sen \cite{aticksen}, 
appear rather convoluted and tedious compared to the pure-spinor 
approach adopted in the present paper. As we shall see in section \ref{53}, our results are in 
agreement with those of \cite{pw}.

Without further ado, let us present here 
the main result of our paper: to all orders in $\alpha^{\prime}$, 
in the four-field approximation, the effective Lagrangian reads
\beal
{\cal L}&=
\frac{1}{2\kappa^2}R-\frac{1}{2}\partial_mD\partial^mD
-\frac{1}{3!}{e^{-\sqrt{2} \kappa D}}H_{mnp}H^{mnp} 
-\frac{1}{\sqrt{2}\kappa }\sum_M\frac{1} {M!}e^{\frac{5-M}{\sqrt{2}} \kappa D}  
\widehat{F}_{m_1\dots m_M}\widehat{F}^{m_1\dots m_M}   \nn\\
&~~~~~~~~~~~~~~~~~~~~~~~~~~~~~~~~~~~~~~~~~~~~~~~~~~~~~~~~~~~~~~~~~~~~~
+\sqrt{2}\sum_{M+N=8}(-1)^{[\frac{M+1}{2}]}\star(B\wedge F_{(M)}\wedge F_{(N)})\nn\\
& +~~\widehat{\mathcal{G}}(\partial, \alpha^{\prime}) \Big\{\frac{1}{4!}
t^{a_1\dots a_8}t_{b_1\dots b_8}\widehat{R}_{a_1a_2}{}^{b_1b_2}\widehat{R}_{a_3a_4}{}^{b_3b_4}
\widehat{R}_{a_5a_6}{}^{b_5b_6}\widehat{R}_{a_7a_8}{}^{b_7b_8}\nn\\
&\qquad\qquad +\sum_{M,N}u_{ij}{}^{mn pq  m'n'p'q'; a_1\dots a_M;b_1\dots b_N}
\widehat{R}_{mnm'n'}
\widehat{R}_{pqp'q'} 
\partial^i{F}_{a_1\dots a_M}\partial^j{F}_{b_1\dots b_N}\nn\\
&\qquad\qquad+\sum_{M,N,P,Q}v^{a_1\dots a_M;b_1\dots b_N;c_1\dots c_P;d_1\dots d_Q}
\partial_{i}\partial_{j}{F}_{a_1\dots a_M}\partial^{i}\partial^{j}{F}_{b_1\dots b_N}
{F}_{c_1\dots c_P}{F}_{d_1\dots d_Q}\Big\}  +{\cal O}({\psi}^2)~,
\label{111}
\end{align}
where $\kappa$ is the gravitational coupling constant, and the action of the operator 
$\widehat{\GV}$ is described below (\ref{thelagrangian}). 
The omitted fermionic terms are given explicitly 
in (\ref{thelagrangian}). 
The tensors $u$, $v$, are defined in appendix \ref{apptraces}; our normalization 
for $t_8$ can be found in section \ref{fpam}. 
The sums over $M,\dots Q$, run over 
even integers from zero to four for IIA supergravity, 
and over odd integers from one to five for IIB. In 
the case of IIB we work in a formalism with an action, 
so that the self-duality of the five-form (at lowest order in $\alpha'$) 
is imposed at the level of the equations of motion. The  
Riemann tensor $\widehat{R}$ is with respect to a modified connection 
with torsion, which includes the NS-NS three-form field and the dilaton, 
see equation (\ref{modifiedconnection}) below. 
In fact, the complete Lagrangian could be written 
in terms of $\widehat{R}$ instead $R$, as the scalar curvature of 
$\widehat{R}$ differs from that of $R$ by a term proportional to 
$\nabla^2 D$, which is a total derivative. 
The RR field strengths $\widehat{F}$ appearing in the 
Lagrangian above obey modified Bianchi identities and are explicitly 
defined in (\ref{modified}). 

The structure of this paper is as follows. In section \ref{review} we give a 
short reminder of the ingredients of the pure-spinor 
string which are used in the present paper. 
In section \ref{norcor} we give the all-order expansion of the vertex operators in flat target space. 
In section \ref{opsamp} we compute the open-string 4-point amplitude and the 
corresponding effective action. In section \ref{closed} we `square' the open-string amplitudes, 
according to the KLT relations, and construct the complete IIA/IIB 
effective action at quartic order in the fields. 
In section \ref{lit} we discuss our results in relation 
to the predictions of the linearized superfield. 
Finally, in section \ref{discussion}, we comment on possible extensions 
and applications of our work. The appendices contain further technical details of the calculation.

\section{Review of the pure-spinor formalism}\label{review}

We now give a brief review of the 
pure-spinor formalism for the superstring, focusing on the points relevant to the present paper. 
The material in this section is well-known; it is included here mainly for the purpose of 
establishing notation/conventions. In the following we are 
adopting conventions as in e.g. \cite{bd}, restoring $\al'$ in addition. 

The pure-spinor string is based on a worldsheet conformal field theory with fields $x^{m},
\theta^{\al}$ corresponding 
to the coordinates of the target superspace, and a (worldsheet scalar, target-space spinor) 
ghost $\lambda^{\al}$, 
plus the conjugate momenta. For the type IIA  (resp. IIB) 
string, we also have the right-movers $\tilde\q^{\bar\al},\tilde \la ^{\bar \al}$, 
with chirality opposite to (resp. same as) that of their left-moving counterparts. 
The worldsheet action in a flat background is given by 
\beal
S=\int {
d^2z\{
-\frac{1}{2}\del x^m\bar{\del}x_m 
-p_\al \bar{\del}\q^\al- \tilde{p}_{\bar\al}\del\tilde{\q}^{\bar\al}
+w_\al\bar{\del}\la^\al 
+\tilde{w}_{\bar\al} \del\tilde{\la}^{\bar\al}
\}
} \,.
\end{align}
This looks like a free action, but the ghosts are not 
really free fields as they are subject to the  pure-spinor constraint
\beal\label{puresp}
\la \g^{m} \la = 0 \,.  
\end{align}
This constraint has several consequences: first, it reduces the number of 
independent components of $\lambda$ from 32 to 22, so that the central charge vanishes and the theory is 
critical in ten dimensions. The stress energy tensor is 
\beal
T&= -\h \del x^m\del x_m -p_\al \del\q^\al +w_\al\del\la^\al \nn\\
\tilde{T}&= -\h \bdel x^m\bdel x_m -\tilde{p}_{\bar\al} \bdel\tilde{\q}^{\bar\al} +\tilde{w}_{\bar\al}
\bdel\tilde{\lambda}^{\bar\al} ~.
\end{align}
Secondly, the constraint implies the presence of a gauge symmetry for the momentum: 
\beal 
\delta w_{\al} = \Lambda^{m} \g_{m} \la \,. 
\end{align} 
The observables of the theory must be gauge-invariant, and one can show that 
they can always be expressed in terms of the currents
\beal
J&:=(w\la)\nn\\
N_{mn}&:=\h (w\g_{mn}\la)~,
\end{align}
which are respectively the ghost-number current and the Lorentz generators in the ghost sector. 
As it turns out, 
one can write down a BRST operator to select the physical states. Let us first make 
the following definitions: 
\beal 
d_\al&:=p_\al-\h (\g^m\q)_\al \del x_m -\frac{1}{8} (\g^m\q)_\al
(\q\g_m\del\q)\nn\\ 
\Pi^m&:=\del x^m+\h(\q\g^m\del\q)~.
\end{align}
%
%Canonical dimensions:
%%
%\beal
%[w_\al] &=M^{-3/2}\nn\\
%[x^m], ~[N_{mn}], ~[J], ~[Q]&= M^{-1}\nn\\
%[\q^\al], ~[p_\al], [d_\al] &= M^{-1/2}\nn\\
%[\Pi^m], ~[{\cal L}]&= M^0\nn\\
%[D_\al], ~[A_\al], ~[\la^\al]&=M^{1/2}\nn\\
%[A_m], ~[\del], ~[\del_m], ~[V], ~[U] &=M^1\nn\\
%[W^\al]&=M^{3/2}\nn\\
%[F_{mn}]&=M^2~.
%\end{align}
%
From the action one can derive the following OPE's 
\beal
x^m(y)x^n(z)&\sim  -\al' \eta^{mn}log\frac{|y-z|^2}{\al'} \, , & 
p_\al(y)\q^\be(z)&\sim   \frac{\al'}{y-z}\delta_\al^\be\nn\\
d_\al(y) d_\be(z)&\sim -\frac{\al'}{y-z}\g^m_{\al\be}\Pi_m  \,, &
d_\al (y)\Pi^m(z)&\sim \frac{\al'}{y-z}(\g^m\del\q)_\al \nn\\
d_\al(y)\del\q^\be(z)&\sim \frac{\al'}{(y-z)^2}\delta_\al^\be  \,, &
\Pi^m(y)\Pi^n(z)&\sim -\frac{\al'}{(y-z)^2}\eta^{mn}~.
%\nn\\
%N_{mn}(y)\la^\al(z)&\sim \frac{\al'}{2(y-z)}(\g_{mn}\la)^\al  \nn\\
%J(y)\la(z)&\sim \frac{\al'}{y-z}\la^\al  \nn\\
%N^{kl}(y)N^{mn}(z)&\sim -\frac{3(\al')^2}{(y-z)^2}\eta^{n[k}\eta^{l]m}
%+\frac{\al'}{y-z}(\eta^{m[l}N^{k]n}-\eta^{n[l}N^{k]m}   )  \nn\\
%J(y)J(z)&\sim -\frac{4(\al')^2}{(y-z)^2}\nn\\
%J(y) N(z)&\sim regular
\end{align}
For any superfield $V(x,\q)$, 
\beal
\Pi^m(y)V(z)&\sim - \frac{\al'}{y-z}\del^m V \,,\nn\\
d_\al(y) V(z)&\sim \frac{\al'}{y-z}D_\al V  ~,
\end{align}
where $D_\al:= \frac{\del}{\del\q^\al}+\h (\g^m\q)_\al \del_m$ is the superderivative. 
The BRST operator is 
\beal
Q:=\oint dz \la^\al d_\al ~.
\end{align}
It can easily be 
checked that $Q^{2} = 0$, 
using the OPE's given above and the constraint (\ref{puresp}). 
The physical states are determined by the cohomology of $Q$. 
For massless states, one writes down the most general form for the vertex operators   
\beal
U&:=\la^\al A_\al(x,\theta) \nn\\
V&:=\del\q^\al A_\al +\Pi^m A_m +d_\al W^\al +\h N^{mn} F_{mn} 
\label{openvertices}
\end{align}
and requires that they satisfy $Q U = 0$, $Q V= \al' \del U$, so 
that $\int V$ is BRST-closed. In the next section we will find the solutions of these conditions. 
When studying closed strings one has to consider also the right-moving sector. The total BRST operator is 
then $Q + \tilde Q$, and the vertex has the form $U_{closed} = \la^{\al} 
\tilde\la^{\bar\al} A_{\al \bar \al}$. Furthermore, 
one can take advantage of the fact that the left- and right-movers are decoupled, 
and the cohomology of the total BRST is simply the tensor product of the 
two sectors. This means that every closed vertex is a linear combination of vertices of a factorized form:
\beal
U_{closed}&:=e^{ik\cdot x} \la^\al A_\al(\q) \tilde{\lambda}^{\bar\be} \tilde{A}_{\bar\be}(\bq)   \nn\\
V_{closed}&:=e^{ik\cdot x} (\del\q^\al A_\al +\Pi^m A_m +d_\al W^\al
+\h N^{mn} F_{mn}) \nn\\ 
&~~\otimes(\bdel\tilde{\q}^{\bar\be} \tilde{A}_{\bar\be} +\tilde{\Pi}^m \tilde{A}_m 
+\tilde{d}_{\bar\be} \tilde{W}^{\bar\be} +\h \tilde{N}^{mn} \tilde{F}_{mn})~.
\label{closedvertices}
\end{align}
Note that the factorization takes place for each given momentum $k$,  
which must take the same value in both sectors.

\subsection*{Tree level amplitudes}

The 
$N$-point tree level scattering amplitude is given by (concentrating on the left-movers)
\beal
\mathcal{A}= \langle U_1(z_1)U_2(z_2)U_3(z_3)\int dz_4 V_4(z_4)\dots  
\int dz_N V_N(z_N)  \rangle ~.
\label{scattering1}
\end{align}
Let us assume that after integrating out all nonzero modes  
the amplitude (\ref{scattering1}) takes the form
\beal
\mathcal{A}= 
\int dz_4 \dots  
\int dz_N \langle \la^\al \la^\be \la^\g f_{\al\be\g}(z_r, k_r, \q)
\rangle ~,  
\label{zero}
\end{align}
where $k_r^m, ~r=1\dots N$ are the momenta and $f_{\al\be\g}$ depends 
only on the zero modes of $\q^\al$. The prescription for integrating over 
the zero modes is
\beal
\mathcal{A}=T_{\rho\sigma\tau}^{\al\be\g}
(\frac{\del}{\del\q}\g^{pmn}\frac{\del}{\del\q})
(\g_p\frac{\del}{\del\q})^\rho
(\g_m\frac{\del}{\del\q})^\sigma
(\g_n\frac{\del}{\del\q})^\tau 
\int dz_4 \dots \int dz_N f_{\al\be\g}(z_r, k_r, \q) ~,
\end{align}
where 
\beal
T_{\rho\sigma\tau}^{\al\be\g}:=
\frac{1}{672}
(
\delta^{(\al}_\rho \delta_\sigma^\be\delta_\tau^{\g)}
-\frac{3}{20} \g_m^{(\al\be}\delta_{(\rho}^{\g)}\g^m_{\sigma\tau)}
)
~.
\end{align}
The tensor $T_{\rho\sigma\tau}^{\al\be\g}$ satisfies 
 $T_{\rho\sigma\tau}^{\al\be\g}\g^m_{\al\be} = 
T_{\rho\sigma\tau}^{\al\be\g} \g_m^{\rho\sigma} =0$ 
and has been normalized so that $T_{\al\be\g}^{\al\be\g}=1$. 

%The closed string graviton $h_{mn}$ (including 
%the dilaton and antisymmetric tensor), gravitini $\psi_m^\al$,
%$\bar{\psi}_m^\al$,  
%and RR field-strength $F^{\al\be}$ are also assumed to be factorized,
%%
%\beal
%h_{mn}:= a_m \bar{a}_n; ~~~~\bar{\psi}_m^{\bar{\al}} :=a_m\bar{\xi}^{\bar{\al}};
%~~~~\psi_m^\al :=\bar{a}_m \xi^\al;  
%~~~~F^{\al\bar{\be}}:=\xi^\al\bar{\xi}^{\bar{\be}} ~.
%\end{align}
%%

\section{$\q$-expansions and the vertex operator}\label{norcor}

In this section we give the details relating to the $\theta$-expansion of the 
closed-string vertex operators for type II superstrings in a flat-space background. 
As was already noted in the introduction, we will show that it is possible to obtain 
an explicit exact result to all orders in $\theta$. Since, as we have seen in the previous section, 
the closed-string vertex operators factorize in flat target-space, cf. eqn (\ref{closedvertices}), 
it suffices to consider the open-string  vertex operator.

Ten-dimensional superYang-Mills admits a formulation 
in superspace in terms of on-shell superfields. 
Let $F$ be the supercurvature two-form corresponding to 
superpotential one-form $A$, $F=DA$. It has been known for some time 
that imposing the constraint $F_{\alpha\beta}=0$ on the spinor-spinor 
components of the supercurvature, leads to the superYang-Mills 
equations of motion \cite{sym}. The physical multiplet consists of the 
gauge boson and the gaugino, which can be identified with the 
$\theta=0$ components of the superfields $A_{m}$, $W^{\alpha}$, respectively. 
On-shell these superfields satisfy 
\beal
2D_{(\al}A_{\be)}&=\g_{\al\be}^mA_m\nn\\
D_\al W^{\be}&=\frac{1}{4}(\g^{mn})_{\al}{}^{\be}F_{mn}~,
\label{go}
\end{align}
where 
\beal
F_{mn}&= 2\del_{[m}A_{n]}\nn\\
W^\al&= \frac{1}{10}(\g^m)^{\al\be}(D_\be A_m-\del_m A_\be)
~
\end{align} 
and $D_{\al}$ is the supercovariant spinor derivative. 
Imposing the gauge (this is the analogue of the 
choice of normal coordinates in the case at hand):
\beal
\q^\al A_\al =0~,
\end{align}
allows us to convert the supercovariant spinor derivative 
into an ordinary one: $\q^\al D_\al =\q^\al\partial/\partial\q^\al$.  
Taking (\ref{go}) into account, 
this leads immediately to the following recursion relations:
\beal
A_\al^{(n)}&=\frac{1}{n+1}(\g^m\q)_\al A_m^{(n-1)}\nn\\
A_m^{(n)}&=\frac{1}{n}(\q\g_mW^{(n-1)})\nn\\
W^{\al (n)}&=-\frac{1}{2n}(\g^{mn}\q)^\al\del_{m} A_{n}^{(n-1)}~.
\label{exp1}
\end{align}
These can be solved to give
\beal
A_m^{(2k)}&=\frac{1}{(2k)!}[{\cal O}^k]_m{}^q a_q\nn\\
A_m^{(2k+1)}&=\frac{1}{(2k+1)!}[{\cal O}^k]_m{}^q (\q\g_q\xi)
~,
\label{exp2}
\end{align}
where 
\beal
[{\cal O}]_m{}^q:=\frac{1}{2}(\q\g_m{}^{qp}\q)\del_p
\end{align}
and we have set 
\beal
a_m:=A_m|; ~~~\xi^\al:=W^\al|; ~~~f_{mn}:=F_{mn}|~.
\end{align}
In the equation above, we use the standard notation according to which $S|$ denotes the 
$\theta=0$ component of the superfield $S$. 
Clearly, (\ref{exp1},\ref{exp2}) completely determine the 
$\q$-expansions of all superfields. 
The first few terms in the expansions read:
\beal 
A_\al^{(1)} = \frac12 (\theta \g^m)_\al a_m \, ,\quad  A_\al^{(2)} = \frac13
(\theta \g^m)_\al &(\theta \g_m \xi) \, ,\quad  
A_\al^{(3)} = \frac{1}{16}(\theta \g^m)_\al (\theta
\g_m{}^{pq} \theta)\del_q a_p  ~ ; \nn \\
A_m^{(1)} = (\theta \g_m \xi) \, ,\quad   A_m^{(2)} = \frac14 (\theta
\g_m{}^{pq}& \theta) \del_q a_p \, , \quad  A_m^{(3)} = \frac{1}{12} (\theta
\g_m{}^{qp} \theta) (\theta \g_q\del_p \xi)~. 
\end{align}
The series in (\ref{exp2}) can be formally summed 
to all orders\footnote{Of course the series terminates at $\theta^{16}$.} in $\theta$ to give
\beal
A_m=[{\rm cosh}\sqrt{{\cal O}}]_m{}^qa_q+[{\cal O}^{-1/2}{\rm sinh}\sqrt{{\cal O}}]_m{}^q(\theta\gamma_q\xi)~.
\end{align}
It is interesting to
 observe the remarkable similarity of the expression above to the $\theta$-expansion 
of the gravitino field strength in eleven-dimensional supergravity, 
equation (106) of \cite{curved}. Similar expressions can readily 
be obtained for $A_{\alpha}$, $W^{\alpha}$. 

These formul\ae {} also 
give the complete expansion of the closed string vertex, when the polarizations of the 
`graviton' $\Theta_{mn}$ (which here  includes 
the dilaton and the antisymmetric tensor), the gravitini $\psi_m^\al$,
$\bar{\psi}_m^{\bar\al}$,  
and the RR bispinor field-strength $\slsh{F}^{\al\bar\be}$ are also assumed to be factorized,
\beal
\Theta_{mn}:= a_m \otimes \tilde{a}_n; ~~~~\bar{\psi}_m^{\bar{\al}} :=i\sqrt{2}a_m\otimes\tilde{\xi}^{\bar{\al}};
~~~~\psi_m^\al :=i\sqrt{2}\xi^\al\otimes \tilde{a}_m;  
~~~~\slsh{F}^{\al\bar{\be}}:=\sqrt{\kappa}\xi^\al\otimes\tilde{\xi}^{\bar{\be}} ~,
\label{facto}
\end{align}
where $\kappa$ is the gravitational coupling constant (normalizations have been chosen 
for later convenience). 
In conclusion: plugging the expressions obtained here 
for the $\theta$-expansions of the various superfields in (\ref{openvertices}, \ref{closedvertices}), 
we obtain the explicit form of the 
string vertices in flat target space to all orders in $\theta$, as advertised.

\section{The open-string amplitude}\label{opsamp}

We now have all the ingredients to compute the amplitudes. 
In order to improve the presentation,  
in the following we will confine ourselves to 
presenting the main results of the computation. 
The interested reader may consult the appendices  
for the omitted technical details. 
The first step is to compute the open-string amplitudes, i.e. 
we consider only the left-moving sector. The $x$-dependent part of the correlator is standard: 
\beal
\langle
e^{ik_1\cdot x(z_1)}e^{ik_2\cdot x(z_2)}e^{ik_3\cdot x(z_3)}
e^{ik_4\cdot x(z_4)} \rangle & =
\prod_{i<j}^4 (z_i-z_j)^{\al' k_i\cdot k_j} \equiv \Pi(z_{ij}) \,, \nn \\
\langle
e^{ik_1\cdot x(z_1)}e^{ik_2\cdot x(z_2)}e^{ik_3\cdot x(z_3)}
:\del x^m (z_4)e^{ik_4\cdot x(z_4)}: \rangle & =
\sum_{i=1}^{3} \frac{i\al' k_i^m}{z_i-z_4} \Pi(z_{ij}) \, ~.
%\prod_{i<j}^4 (z_i-z_j)^{\al' k_i\cdot k_j}
\label{xcorrs}
\end{align}

\subsection{The 3-point amplitude}

The computation of the 3-point amplitude is straightforward, since it only contains unintegrated vertices. 
We have 
\beal
\mathcal{A}_3=\langle U_1(z_1)U_2(z_2)U_3(z_3)  \rangle ~.
\end{align}
For three massless particles on-shell, $k_{i} \cdot k_{j} =0$. We find 
\beal\label{3ptop}
{\cal A}_{3}^{{\rm op}}(k_i;a_i,\xi_i)
=  \frac{1}{5760} \Big[ & k_{1} \cdot a_{3} \, a_{1} \cdot a_{2} + k_{2} \cdot a_{1} \, a_{2} \cdot a_{3}
+k_{3} \cdot a_{2} \, a_{1} \cdot a_{3}\nn\\
& -  i \xi_{2} \slsh{a_{1}} \xi_{3}+ i \xi_{1} \slsh{a_{2}} \xi_{3}- i \xi_{1} \slsh{a_{3}} \xi_{2} \Big] \,.
\end{align}

%\beal
%\mathcal{A}_3= \frac{1}{5760} \Big[i  k_{1} \cdot a_{3} \, a_{1} \cdot a_{2} - i k_{1} \cdot a_{2} \, a_{1} \cdot a_{3} 
%+ 2 \xi_{2} \slsh{a_{1}} \xi_{3} \Big] - (2,1,3) - (3,2,1) \,.
%\end{align}

\subsection{The 4-point amplitude}

Here we compute
\beal\label{A4}
\mathcal{A}_4=\langle U_1(z_1)U_2(z_2)U_3(z_3)\int d^2z_4 V_4(z_4) \rangle ~.
\end{align}
This amplitude receives several contributions, 
coming from the different terms in $V_4$ and from the expansion of all the fields in powers of $\theta$. 
For each of these terms, 
the function of the zero modes $f_{\al\be\g}$ in (\ref{zero}) has  
a different structure. 
The detailed computation of all the terms in the 4-point amplitude can be found 
in appendix \ref{amplitudes}. We give here the final result: 
%%
%\begin{center}
%\fbox{\parbox{15cm}{
%
%%%%%
%
\beal
\frac{\al'}{5760} \Big[ \frac{\Pi(z_{ij})}{z_{1}-z_{4}} & \Big\{\,\,\,
2  k_{1} \cdot a_{4} \, k_{2} \cdot a_{3} \, a_{1} \cdot a_{2} 
+2 k_{3}  \cdot a_{4} \, k_{2} \cdot a_{1} \, a_{2} \cdot a_{3} 
+2 k_{1} \cdot a_{3} \, k_{3} \cdot a_{2} \, a_{1} \cdot a_{4} \nn\\
&-2 k_{4} \cdot a_{1} \, k_{2} \cdot a_{3} \, a_{2} \cdot a_{4}   
+ k_{1} \cdot k_{4} \, a_{1} \cdot a_{2} \, a_{3} \cdot a_{4} 
- k_{2}\cdot k_{4} \, a_{2} \cdot a_{3} \, a_{1} \cdot a_{4} \nn\\
&+ 2i\xi_{1} \slsh{a_{2}} \xi_{4} \, k_{2} \cdot a_{3}  \, 
	+  2i\xi_{2} \slsh{a_{3}} \xi_{4} \, k_{4} \cdot a_{1}  \,
	+  i \xi_{1} \slsh{k_{3}} \xi_{4}a_{2} \cdot a_{3}   
	+ i\xi_{3} \slsh{a_{2}} \slsh{a_{1}} \slsh{k_{1}} \xi_{4}\nn\\
	&+i\xi_{2} \slsh{a_{1}} \xi_{3} \, k_{1} \cdot a_{4}  \, 
-i\xi_{2} \slsh{a_{4}} \xi_{3} \, k_{4} \cdot a_{1}  \, 
	+  2i\xi_{1} \slsh{a_{3}} \xi_{2} \, k_{1} \cdot a_{4}  \,
	+   i\xi_{2} \slsh{k_{4}} \xi_{3}a_{1} \cdot a_{4}  \nn\\
 &- i\xi_{3} \slsh{a_{2}} \slsh{a_{4}} \slsh{k_{4}} \xi_{1} 
-(\xi_1\gamma^m\xi_4)(\xi_2\gamma_m\xi_3)
\Big\} \Big] +{\rm cyclic~permutations.}\nn
\end{align}
%
%%%%%
%
%}}
%\end{center}
%%
%\beal\end{align}
%
%
After doing the $z_{4}$ integration (see appendix \ref{integrals}) and 
then summing $(2\leftrightarrow 3)$ we get 
\beal\label{completeamp}
\mathcal{A}_{4}= \frac{2\al^{\prime 2}}{5760} \Big\{&-2 k_{1}\cdot k_{3} \, k_{1}\cdot a_{4} \, k_{2} \cdot a_{3} \, a_{1} \cdot a_{2} -\frac12 k_{1}\cdot k_{4} \, k_{2}\cdot k_{4}\,  a_{1}\cdot a_{2} \, a_{3}\cdot a_{4}\nn\\
&-ik_{1}\cdot k_{3}\Big[
 a_{2}\cdot a_{3}\, (\xi_1\slsh{k_3}\xi_4) 
-k_{3}\cdot a_{2}\, (\xi_1\slsh{a_3}\xi_4)
+ k_{2}\cdot a_{3}\, (\xi_1\slsh{a_2}\xi_4)
\Big]\nn\\
&+ik_{1}\cdot k_{4}\Big[
 k_{1}\cdot a_{3}\, (\xi_1\slsh{a_2}\xi_4)
+\frac{1}{2}(\xi_1\slsh{a_3}\slsh{k_3}\slsh{a_2}\xi_4)
\Big]\nn\\
&-\frac{1}{3} k_{1}\cdot k_{2} ~(\xi_1\g_m\xi_4)(\xi_2\g^m\xi_3) 
\Big\} + {\rm permutations}  \,. 
\end{align}

\subsection{The open-string effective action}

The action which reproduces the amplitude (\ref{completeamp}) is 
\beal
S_{2} \propto \al^{\prime 2} 
%\2\times 5760
\int d^{10}x \, \GV(\partial, \al') \, 
\Big\{ &\tr f^{4} - \frac14 (\tr f^{2})^{2}
-4if_{mn}f_{mp}(\xi\g^n\partial^p\xi)-2if_{mn}f_{pq}(\xi\g^{mnp}\partial^q\xi)\nn\\
&+\frac{4}{3}(\xi\g_m\partial_n\xi)(\xi\g^m\partial^n\xi)  \Big\}~,
\label{drept}
\end{align}
where the operator $\GV(\partial, \al')$
 should  be understood as follows: one has to split the positions of the different 
fields, take the Fourier transform and insert $\GV(k_{i})$, which 
is defined in appendix \ref{integrals}.

\section{The closed-string amplitude}\label{closed}

The closed-string amplitude can be readily reconstructed using the formul{\ae} of \cite{klt}. 
In particular, denoting by ${\cal A}^{op}_{N}$ the N-point open string amplitude, the N-point closed string amplitude reads:
\beal\label{klt}
{\cal A}^{\rm cl}_{N} = & (\frac{i}{2})^{N-3}  g^{N-2} \sum_{P,P'}{\cal A}^{\rm op}_{N} (P) 
\otimes{\cal{ \tilde A}}^{\rm op}_{N}(P') e^{i \pi f(P,P')} \, .
\end{align}
Here the sum is over different orderings $P,P'$ of the open string vertices. 
In the case of 3 and 4 points the sum actually consists of only one term. We will need the explicit expression for these cases: 
\beal\label{klt3}
{\cal A}^{\rm cl}_{3} = &  g \, {\cal A}^{\rm op}_{3} \otimes{\cal{ \tilde A}}^{\rm op}_{3} \, 
\end{align}
and
\beal\label{klt4}
{\cal A}^{\rm cl}_{4} = & -   g^2 \sin(\pi \al' k_{2}\cdot k_{3}){\cal A}^{\rm op}_{4} (\al' s/2,\al' t/2) 
\otimes{\cal{ \tilde A}}^{\rm op}_{4}(\al' t/2,\al' u/2) \, .
\end{align}
Here $s,t,u$ are the usual Mandelstam variables for 4-particle scattering: 
\begin{align}
s= - 2 k_{12}= - 2 k_{34} \, ;  t = -  2 k_{14}= - 2 k_{23} \,; u =- 2 k_{13}= - 2 k_{24} \,. \nn
\end{align}

 The constant $g$ in (\ref{klt3}),(\ref{klt4}) is the closed string coupling, and the amplitudes on the right hand side should be taken without the corresponding open string coupling. It follows in particular that the normalization of the open amplitudes is irrelevant for this computation, since it can be changed by redefining the open string coupling.

\subsection{The 3-point amplitude}

At the level of 3-point amplitudes, the normalization can be determined 
by matching with the coefficients of the quadratic effective action. 
For higher-point functions, the requirement of unitarity 
is sufficient to fix the normalization; for instance, 
the 4-point function will have poles that must come from 1-particle-reducible diagrams, 
so their coefficient is determined by the 3-point amplitudes. 
All this is well-known and for the purely-gravitational part of 
the effective action the computations have been 
explained in detail by Gross and Sloan \cite{grosssloan}. 
We follow their conventions and normalizations in this section. 

Using (\ref{3ptop}) and (\ref{klt3}) we find for the bosonic part:
\beal
{\cal A}^{\rm cl}_{3b} & =  g \Big( k_2^mk_2^n\Theta_{1,mn}\Theta_{2,pq}\Theta_3^{pq}
+k_3^mk_2^q\Theta_{2,mn}\Theta_{3,pn}\Theta_{1,pq}
+k_1^mk_2^q\Theta_{3,mn}\Theta_{2,pn}\Theta_{1,pq}\nn\\
&+ \frac{1}{\kappa} \slsh{F}_{1}^{\al \bar\al} \slsh{F}_{2}^{\beta \bar\beta} 
\Theta_{3, mn} \g^{m}_{\al \beta}\g^{n}_{\bar\al \bar\beta}
+{\rm cyclic}\Big)\, , 
\label{iuo}
\end{align}
where $\Theta_{mn}:=a_m\otimes \tilde{a}_n$,  
$\slsh{F}^{\al \bar\al}:= \sqrt{\kappa} \, \xi^{\al}\otimes\tilde{\xi}^{\bar\al}$, and $g=2 \kappa$. 
Moreover in our conventions 
the symmetric, traceless part of $\Theta_{mn}$ is given by 
$h_{mn}$,  $g_{mn}=\eta_{mn}+2\kappa h_{mn}$; the antisymmetric part is $B_{mn}$, 
$H_{mnp}=3\partial_{[m}B_{np]}$; the trace part is set equal to 
$(\eta_{mn}-k_m\tilde{k}_n-\tilde{k}_mk_n)D/\sqrt{8}$.

The fermionic part of the amplitude can be written as
\beal
{\cal A}^{\rm cl}_{3f} & = - \frac{g}{2} \Big(
-ik_1^{[m}h_1^{n]p}\psi_{2m}\g_p\psi_{3n}-\frac{i}{2}k_1^nh_{3np}\psi_{1m}\g^p\psi_2^m + (\psi\rightarrow \bar\psi ) \nn\\
&-\frac{1}{\sqrt{\kappa}}\psi_{2m}\g^n\slsh{F}_3\g^m\bar{\psi}_{1n}
+{\rm permutations}\Big)\, .
\label{iuob}
\end{align}

\subsection*{Fermionic terms}

The fermionic field $\bar \psi_{m}^{\bar \alpha} := i \sqrt{2} a_{m} \otimes \tilde \xi^{\bar \al}$,   
can be decomposed on-shell into the gamma-traceless (spin 3/2) and gamma (spin 1/2) parts: 
$\psi_{m } = \chi_{m} + \frac{i}{\sqrt{8}} \g_{m} \lambda$. 
In terms of the $\chi$, $\lambda$ fields, the fermionic kinetic terms read 
\begin{align} 
\frac12  \chi_{m} \gamma^{mnp} D_{n}  \chi_{p}  
+\frac12  \lambda \g^{m} D_{m}  \lambda + (\chi \rightarrow \bar\chi, \la \rightarrow \bar\la) \,.
\label{kikin}
\end{align}
In this equation, $D_{m}\chi_{n} = \partial_{m} \chi_{n} - \frac14 \omega_{m}^{~ab} \g_{ab}\chi_{n}$, 
and the linearized spin connection is given by 
$\omega_{m ab} = 2 \kappa \partial_{[a} h_{b]m}$. 
%From the kinetic term, one finds a 3 point coupling with the graviton 
%\beal
% -\frac{\kappa}{8} \partial_{m} h_{np} \bar\chi^{m} \g^{p} \bar\chi^{n} 
% \end{align}
%
On the other hand, the fermionic 3-point amplitude (\ref{iuob}) corresponds to the Lagrangian
\beal
& -\kappa \big( \frac12 \Theta_{np}  \psi_{m} \g^{p} \partial^{n} \psi^{m} 
+ \partial_{m} \Theta_{n p}\psi^{m} \g^{p} \psi^{n} \big) + (\psi\rightarrow \bar\psi )\nn \\
& + \sqrt{\kappa}  \psi_{m} \g^{p} \slsh{F} \g^{m} \bar{\psi}_{p} \,.
\label{firfiriki}
\end{align}
When $\Theta$ is the graviton, it can be seen that 
these couplings are accounted for by  the three-point contribution 
coming from the kinetic term (\ref{kikin}). 
The remaining contribution 
(obtained by letting $\Theta$ be the dilaton or the antisymmetric tensor) 
cannot come from the kinetic term. We thereby obtain the 
fermionic 3-point Lagrangian (completed as usual by the appropriate dilaton couplings)
\begin{align}
&\frac{\kappa}{24} 
e^{-\frac{\kappa D }{\sqrt{2}} } 
 H^{m_1m_2m_3}( \chi_{n}\g^{[n} \g_{m_1m_2m_3}\g^{p]} \chi_{p} )+ (\chi\rightarrow \bar\chi ) \nn\\
& -2 \sqrt{\kappa}
\sum_p \frac{c_p}{p!} e^{\frac{5-p}{2 \sqrt{2}} \kappa D } F^{m_1\dots m_p}  (\chi_{n} \g^{[n} 
\gamma_{m_1\dots m_p} \g^{p]} \bar{\chi}_{p})  +\dots~,
\label{nnm}
\end{align}
%
%- \frac{5 i \kappa}{4} \partial_{m} D \chi^{m} \lambda +
%
%
where we have expanded
\beal
\slsh{F}^{\al \bar\al}=\sum_p \frac{c_p}{p!}(\gamma_{m_1\dots m_p})^{\al \bar\al}F^{m_1\dots m_p}; ~~~~~
c_p^2=\frac{(-1)^{p+1}}{16\sqrt{2}}~.
\label{fexp}
\end{align}
The ellipses in equation 
(\ref{nnm}) signify ${\cal O}(\lambda^2)$, as well as ${\cal O}(\chi\lambda)$ 
terms, which we have 
omitted for simplicity. The interested 
reader can find the complete fermionic Lagrangian in the literature
\footnote{To compare, for example, with the conventions of  
\cite{giani}, one should  set $\sqrt{2}\kappa\rightarrow 1$ in the 
formul\ae {} of the present paper,  and 
make the following substitutions 
in equation (1.19) of that reference: 
$\phi\rightarrow D/\sqrt{2}$ ~(i.e. $\sigma\rightarrow e^{D/4}$), 
$\psi\rightarrow i\sqrt{2}\chi$, $\lambda\rightarrow i\sqrt{2}\lambda$, 
 with all the 
remaining fields unchanged.}. 
In order to bring the three-point couplings in the form above, we
 added to (\ref{firfiriki}) certain terms which vanish on-shell.

\subsection*{Bosonic terms}

The bosonic amplitude (\ref{iuo}) corresponds to an effective 
3-point Lagrangian
\beal\label{lag3pt}
{\cal L}_{3}=\frac{1}{2\kappa^2}R-\frac{1}{2}\partial_mD\partial^mD
&-\frac{1}{3!}{e^{-\sqrt{2} \kappa D}}H_{mnp}H^{mnp} 
-\frac{1}{\sqrt{2}\kappa }\sum_p\frac{1} {p!}e^{\frac{5-p}{\sqrt{2}} \kappa D}  
F_{m_1\dots m_p}F^{m_1\dots m_p}  \nn\\
& +\sqrt{2}\sum_{p+q=8}(-1)^{[\frac{p+1}{2}]}\star(B\wedge F_{(p)}\wedge F_{(q)})
- \sqrt{2} \sum_{p} \frac1{p!} F_{j_{1}\dots j_{p}} B_{mn} F^{j_{1}\dots j_{p}mn} \,.
\end{align}
For type IIB, we are using a formalism with an action, so that the self-duality of the five-form 
(at lowest-order in $\alpha'$) is imposed at the level of the equations of motion. 
The last term of (\ref{lag3pt}) contains couplings which depend on the NS and RR potentials, 
rather than just on the field strengths. 
As is well-known, they can be reabsorbed in the kinetic term for the RR fields if 
one introduces the modified field strengths 
\beal
\widehat{F}_{(p)} & = F_{(p)} + 2 \kappa~(-1)^{p} \, C_{(p-3)} \wedge H~;~~~~~p\geq 3 ~. 
\label{modified}
\end{align}

\subsection{The 4-point amplitude}
\label{fpam}

Up to an irrelevant normalization, 
the 4-point open-string amplitude in the $u-s$ channel can be written as:
\beal
{\cal A}_4^{{\rm op}}(u,s)=K_{SS}(k_i,a_i) G(u,s)~,
\end{align}
where
\beal
K_{SS}(k_i;a_i,\xi_i) :=8\al^{\prime 2}
\Big\{&2 k_{1}\cdot k_{3} \, k_{1}\cdot a_{4} \, k_{2} \cdot a_{3} \, a_{1} \cdot a_{2} 
+\frac12 k_{1}\cdot k_{4} \, k_{2}\cdot k_{4}\,  a_{1}\cdot a_{2} \, a_{3}\cdot a_{4}\nn\\
&+ik_{1}\cdot k_{3}\Big[
 a_{2}\cdot a_{3}\, (\xi_1\slsh{k_3}\xi_4) 
-k_{3}\cdot a_{2}\, (\xi_1\slsh{a_3}\xi_4)
+ k_{2}\cdot a_{3}\, (\xi_1\slsh{a_2}\xi_4)
\Big]\nn\\
&-ik_{1}\cdot k_{4}\Big[
 k_{1}\cdot a_{3}\, (\xi_1\slsh{a_2}\xi_4)
+\frac{1}{2}(\xi_1\slsh{a_3}\slsh{k_3}\slsh{a_2}\xi_4)
\Big]\nn\\
&+\frac{1}{3} k_{1}\cdot k_{2} ~(\xi_1\g_m\xi_4)(\xi_2\g^m\xi_3) 
\Big\} + {\rm permutations}  \,, 
\label{59}
\end{align}
and $G(u,s)$ is defined in appendix \ref{integrals}. 
It is useful to note that 
\beal 
t_{abcdefgh} M_{1}^{ab} M_{2}^{cd} M_{3}^{ef} M_{4}^{gh} = & -2\, 
(\tr M_{1}M_{2} \tr M_{3}M_{4} + \tr M_{1}M_{3} \tr M_{2}M_{4} + \tr M_{1}M_{4} \tr M_{2}M_{3}) \nn\\
& + 8 \, (\tr M_{1}M_{2}M_{3}M_{4} + \tr M_{1}M_{3}M_{2}M_{4} + \tr M_{1}M_{3}M_{4}M_{2}) \, .
\end{align}
In particular for $f^{mn}_i:=2k^{[m}_ia^{n]}_i$, we find
\beal
t_{mnpqrstu}f^{mn}_1f^{pq}_2f^{rs}_3f^{tu}_4=
&-8 k_{1}\cdot k_{3} \, k_{1}\cdot a_{4} \, k_{2} \cdot a_{3} \, a_{1} \cdot a_{2} \nn\\
&-2 k_{1}\cdot k_{4} \, k_{2}\cdot k_{4}\,  a_{1}\cdot a_{2} \, a_{3}\cdot a_{4}
+{\rm permutations}~,
\end{align}
so that the purely bosonic part of (\ref{59}) is equal to $-2\alpha^{\prime2}t_8f^4$.
Using the relations (\ref{klt4}), we write for the closed 4-point amplitude
\beal
{\cal A}^{cl}_{4} = N f(s,t,u) K_{SS} \otimes\tilde K_{SS}~,
\end{align}
where $N$ is a normalization factor to be fixed, and 
\beal\label{f}
f(s,t,u) = & {\rm sin}(-\pi \al' \frac{t}{2}) 
{ G}(\frac{\al' s}{2},\frac{\al' t}{2}){ G}(\frac{\al' t}{2},\frac{\al' u}{2}) \nn\\
& \sim - \frac{8 \pi}{\al^{\prime3} s t u} -2\pi  \zeta(3) +{\cal O}(\alpha^{\prime2}) \,.
\end{align}

\subsection*{Proof of factorization}

Let us now come to the proof of equation (\ref{tensor}) mentioned already 
in the introduction. It is easy to see, concentrating for simplicity 
on the kinematic part and  dropping the operator $\GV$, 
that the 4-point open-string amplitude is of the form
$$
{\cal A}^{op}_{4}\sim K_{SS} \sim \Big( \widehat{{\cal L}}^{op}_4+{\rm permutations} \Big)~,
$$
where $\widehat{{\cal L}}^{op}_4$ is obtained from the 4-point SYM 
Lagrangian (\ref{drept}) by splitting the positions of the fields and 
taking the Fourier transform.  On the other hand, we can write 
$$
K_{SS} =k_{SS}+{\rm permutations}~,
$$
for some $k_{SS}(k_i;a_i,\xi_i)$, so that the permutations act 
on the positions of the particles. By comparing the two expressions it is clear 
that  $\widehat{{\cal L}}^{op}_4$ 
can be identified with $k_{SS}$. 
It follows that the closed-string amplitude is of the form
\beal
{\cal A}^{cl}_{4}&\sim \Big(k_{SS}+{\rm permutations}  \Big)
\otimes \Big(\tilde{k}_{SS}+{\rm permutations} \Big)\nn\\
&= K_{SS}\otimes \tilde{k}_{SS}+{\rm permutations}~, \nn
\end{align}
where the permutations in the second line above act `diagonally', i.e. they act 
simultaneously on both the left an the right sectors. It follows that 
$K_{SS}\otimes \tilde{k}_{SS}$ is obtained from the closed 
4-point Lagrangian by splitting the positions of the fields and 
taking the Fourier transform. Recalling the relation of 
$\tilde{k}_{SS}$ to the 4-point open Lagrangian, we finally arrive at
$$
{\cal L}_4^{cl}={\cal L}_4^{op}\otimes \tilde{{\cal L}}_4^{op}~,
$$
where the action of $\otimes$ should be understood as taking all pairs of 
fields formed by one field in the left and one field in 
the right sector, and using the 
(Fourier transform of the) 
factorization formul\ae {} (\ref{facto}) to convert each  pair 
of open fields to a closed field.

\subsection{The closed-string effective action}
\label{53}

$\bullet$ {NS-NS}

As was observed in \cite{grosssloan},  we can set 
$f_{ab}\otimes\tilde f_{cd} = \frac{1}{\kappa} \widehat{R}_{abcd}$, 
where  we have introduced a modified Riemann tensor
\beal
\widehat{R}_{mn}{}^{pq}:=R_{mn}{}^{pq}
+2\kappa e^{-\frac{\kappa D}{\sqrt{2}}}\nabla_{[m}H_{n]}{}^{pq}
-\sqrt{2}\kappa\delta_{[m}{}^{[p}\nabla_{n]}\nabla^{q]}D
~.
\label{modifiedconnection}
\end{align}
The modified Riemann tensor can be thought of as corresponding  
to a connection with torsion. 
At the linearized level it 
obeys the following identities 
\beal 
\nabla_{[m}\widehat{R}_{np]qr}&=\nabla_{[m|}\widehat{R}_{np|qr]}=0\nn\\
\nabla^i\widehat{R}_{imnp}&=\nabla^i\widehat{R}_{mnip}=0\nn\\
\widehat{R}_{[mnp]q}&=\frac{2\kappa}{3}\nabla_qH_{mnp}\nn\\
\widehat{R}_{mnpq}(H)&=\widehat{R}_{pqmn}(-H)~,  
\label{reqs}
\end{align}
which will be used in the following. 

The NS-NS part of the bosonic effective action comes from the terms of 
the form $f^{4}\otimes\tilde{f}^{4}$ in the closed four-point amplitude. 
We thus find 
\beal
{\cal L}_{\tiny{NS-NS}}= - \frac{(\al')^{3}}{2^{7}4! \pi \kappa^{2}} f(s,t,u) 
t_8t_8 \widehat{R}^4
 ~.
\label{pipa}
\end{align}

$\bullet$ {$(\partial F)^2R^2$}

The two-graviton/two-RR part of the effective action, comes from cross terms 
of the form $f\xi^2\otimes \tilde{f}\tilde{\xi}^2$ 
in the closed four-point amplitude. We thus find
\beal
{\cal L}_{\tiny{(\partial F)^2R^2}}= \frac{(\al')^{3}}{2 \pi \kappa} f(s,t,u) 
 (A_1+\frac{1}{2}A_2+\frac{1}{4}A_3)
 ~.
\label{pipb}
\end{align}
In the equation above we have defined
\beal
A_1&:=\widehat{R}^i{}_{n}{}^j{}_{n'}\widehat{R}_{ipjp'}
<\g^n\partial^p\slsh{F}\g^{(n'}\partial^{p')}\slsh{F}^{Tr}>\nn\\
A_2&:=\widehat{R}_{mn}{}^i{}_{n'}\widehat{R}_{pqip'}\Big(
<\g^{mnp}\partial^q\slsh{F}\g^{(n'}\partial^{p')}\slsh{F}^{Tr}>
+<\g^{mnp}\partial^q\slsh{F}^{Tr}\g^{(n'}\partial^{p')}\slsh{F}>\Big)\nn\\
A_3&:=\widehat{R}_{mnm'n'}\widehat{R}_{pqp'q'}
<\g^{[mnp}\partial^{q]}\slsh{F}\g^{m'n'p'}\partial^{q'}\slsh{F}^{Tr}>~,
%A_3&:=\widehat{R}_{mnn'}{}^i\widehat{R}_{pip'q'}\Big(
%3<\g^{mnp}\partial^{[n'}\slsh{F}\g^{p'}\partial^{q']}\slsh{F}^{Tr}>
%+\frac{1}{2}<\g^{mnp}\partial^j\slsh{F}\g^{n'p'q'}\partial_{j}\slsh{F}^{Tr}>
%\Big)
\end{align}
where the notation $<\dots >$ denotes the trace in the spinor indices. 
In order to bring (\ref{pipb}) to this form, one has to perform an integration by parts,
making use of the Bianchi identities and of the 
equations of motion (\ref{reqs}), which leads to the following relations: 
\beal
A_1&=-\widehat{R}^i{}_{n}{}^j{}_{n'}\widehat{R}_{ipjp'}
<\slsh{F}\g^{(n'}\partial^{p')}\partial^p\slsh{F}^{Tr}\g^{n}>
\nn\\
A_2&=\widehat{R}_{mn}{}^i{}_{n'}\widehat{R}_{pqip'}\Big(
<\slsh{F}^{Tr}\g^{(n'}\partial^{p')}\partial^q\slsh{F}\g^{mnp}>+<\slsh{F}\g^{(n'}\partial^{p')}\partial^q\slsh{F}^{Tr}\g^{mnp}>
\Big)
\label{fddfa}
\end{align}
and
\beal
\widehat{R}_{mnm'n'}\widehat{R}_{pqp'q'}&\Big(
<\g^{[mnp}\partial^{q]}\slsh{F}\g^{m'n'p'}\partial^{q'}\slsh{F}^{Tr}>
+<\slsh{F}\g^{[m'n'p'}\partial^{q']}\partial^{q}\slsh{F}^{Tr}\g^{mnp}>
\Big)\nn\\
&=\widehat{R}_{mnn'}{}^i\widehat{R}_{pip'q'}\Big(
3<\g^{mnp}\partial^{[n'}\slsh{F}\g^{p'}\partial^{q']}\slsh{F}^{Tr}>
+\frac{1}{2}<\g^{mnp}\partial^j\slsh{F}\g^{n'p'q'}\partial_{j}\slsh{F}^{Tr}>
\Big)~.
\end{align}
In addition, in the linearized approximation around flat space we have 
$R_{mn}{}^{pq}\sim \partial_{[m}\partial^{[p}h_{n]}{}^{q]}$. Taking this into 
account, one can prove the following identity
\beal
\widehat{R}_{mnn'}{}^i\widehat{R}_{pip'q'}\Big(
3<\g^{mnp}\partial^{[n'}\slsh{F}\g^{p'}\partial^{q']}\slsh{F}^{Tr}>
+\frac{1}{2}<\g^{mnp}\partial^j\slsh{F}\g^{n'p'q'}\partial_{j}\slsh{F}^{Tr}>\Big)
=2A_3~,
\end{align}
or, equivalently,
\beal
A_3=\widehat{R}_{mnm'n'}\widehat{R}_{pqp'q'}
<\slsh{F}\g^{[m'n'p'}\partial^{q']}\partial^{q}\slsh{F}^{Tr}\g^{mnp}>~.
\label{fddfb}
\end{align}
Putting all the pieces together, we arrive at (\ref{pipb}).

We can now compare our results to the corresponding ones {} 
in \cite{pw}. Indeed, equation (2.13) of that reference 
exactly reproduces equation (\ref{pipb}) of the present paper. 
Note that had the authors of \cite{pw} made a 
different choice of picture changing insertions, they 
would have instead arrived at terms of the form 
$F \partial^{2} FR^2$, as on the right-hand sides 
of equations (\ref{fddfa}, \ref{fddfb}). 
In other words, these two equations can be proven on-shell 
in the linearized approximation, by virtue of the 
picture-changing independence. 
In appendix \ref{last} we shall give a brute-force derivation of 
(\ref{fddfb}) in the case of the $(\partial F_{(1)})^2R^2$ couplings.

$\bullet$ {$(\partial F)^4$}

The purely RR part of the effective action comes from tensoring 
two copies of the purely-fermionic part of the open-string amplitude. 
We have
\beal
{\cal L}_{\tiny{(\partial F)^4}}= -\frac{(\al')^{3}}{36 \pi } f(s,t,u) 
(B_{1}-5 B_{2}+ B_{3}+4 B_{4}- B_{5})
 ~,
\label{pipc}
\end{align}
where we have defined
\beal
B_{1}&:=<\partial_m\partial_p\slsh{F}\g_q\partial^m\partial^p\slsh{F}^{Tr}\g_n\slsh{F}\g^q\slsh{F}^{Tr}\g^n>\nn\\
B_{2}&:=<\partial_m\partial_p\slsh{F}\g_q\slsh{F}^{Tr}\g_n\partial^m\partial^p\slsh{F}\g^q\slsh{F}^{Tr}\g^n>\nn\\
B_{3}&:=<\partial_m\partial_p\slsh{F}\g_q\slsh{F}^{Tr}\g_n\slsh{F}\g^q\partial^m\partial^p\slsh{F}^{Tr}\g^n>\nn\\
B_{4}&:=<\partial_m\partial_p\slsh{F}\g_q\slsh{F}^{Tr}\g_n><\partial^m\partial^p\slsh{F}\g^q\slsh{F}^{Tr}\g^n>\nn\\
B_{5}&:=<\slsh{F}\g_q\slsh{F}^{Tr}\g_n><\partial^m\partial^p\slsh{F}\g^q\partial_m\partial_p\slsh{F}^{Tr}\g^n> \,.\nn\\
\end{align}
In order to bring (\ref{pipc}) to this form, we have integrated by parts, taking the 
linearized Bianchi identities and equations of motion into account, to arrive at the 
following relations:
\beal
<\partial_m\partial_p\slsh{F}\g_q\slsh{F}^{Tr}\g_n><\partial^m\slsh{F}\g^q\partial^p\slsh{F}^{Tr}\g^n>&=-B_{4}+\frac{1}{2}B_{5}\nn\\
<\partial_m\slsh{F}\g_q\partial_p\slsh{F}^{Tr}\g_n><\partial^m\slsh{F}\g^q\partial^p\slsh{F}^{Tr}\g^n>&=B_{4}\nn\\
<\partial_m\partial_p\slsh{F}\g_q\partial^m\slsh{F}^{Tr}\g_n\partial^p\slsh{F}\g^q\slsh{F}^{Tr}\g^n>&=\frac{1}{2}(-B_{1}-B_{2}+B_{3})\nn\\
<\partial_m\partial_p\slsh{F}\g_q\partial^m\slsh{F}^{Tr}\g_n\slsh{F}\g^q\partial^p\slsh{F}^{Tr}\g^n>&=\frac{1}{2}(-B_{1}+B_{2}-B_{3})\nn\\
<\partial_m\partial_p\slsh{F}\g_q\slsh{F}^{Tr}\g_n\partial^m\slsh{F}\g^q\partial^p\slsh{F}^{Tr}\g^n>&=\frac{1}{2}(B_{1}-B_{2}-B_{3})\nn\\
<\partial_m\slsh{F}\g_q\partial_p\slsh{F}^{Tr}\g_n\partial^m\slsh{F}\g^q\partial^p\slsh{F}^{Tr}\g^n>&=B_{2}
~.
\end{align}

$\bullet$ $ \psi\partial\psi (\partial F)^2$

These terms come from tensoring a copy of the  
purely-fermionic part  of the 
open-string amplitude,  
with a copy of the quadratic-fermion part. We thus find 
\beal
{\cal L}_{\psi\partial\psi (\partial F)^2}=
 &\frac{i(\al')^{3}}{6\pi } f(s,t,u) \Big\{
(\psi_{nk}\g_i\partial_j\psi^k{}_{p})<\g^i\partial^j\slsh{F}\g^{(n}\partial^{p)}\slsh{F}^{Tr}>
+(\psi_{nk}\g^i\partial^j\slsh{F}\g^{(n}\partial^{p)}\slsh{F}^{Tr}\g_i\partial_j\psi^k{}_{p})
\nn\\
+&\frac{1}{2 } 
(\psi_{mn}\g_i\partial_j\psi_{pq})<\g^i\partial^j\slsh{F}\g^{[mnp}\partial^{q]}\slsh{F}^{Tr}>
+\frac{1}{2 } (\psi_{mn}\g^i\partial^j\slsh{F}\g^{[mnp}\partial^{q]}\slsh{F}^{Tr}\g_i\partial_j\psi_{pq})
\Big\}
~.
\end{align}

\vfill\break

$\bullet$ $ \psi\partial\psi R^2$

These terms come from tensoring a copy of the purely-bosonic part of the 
open-string amplitude   
with a copy of the quadratic-fermion part. We thus find 
\beal
{\cal L}_{\psi\partial\psi R^2}= \frac{i(\al')^{3}}{64\pi } f(s,t,u) t_8^{a_1\dots a_8} \Big\{
R_{a_1a_2n}{}^i&R_{a_3a_4pi} (\psi_{a_5a_6}\g^{n}\partial^{p}\psi_{a_7a_8})\nn\\
&+\frac12 R_{a_1a_2mn}R_{a_3a_4pq} (\psi_{a_5a_6}\g^{mnp}\partial^{q}\psi_{a_7a_8})
\Big\}~.
\end{align}

$\bullet$ $ \partial^2\psi^4 $

These terms come from tensoring a copy of the purely-bosonic
 part  of the 
open-string amplitude,  
with a copy of purely-fermionic part, or from $f^2 \xi^2  \otimes \tilde f^2  \tilde \xi^2$. We thus find 
\beal
{\cal L}_{\partial^2\psi^4 }&=- \frac{(\al')^{3}}{16\pi } f(s,t,u) \times \nn\\ 
& \Big\{ 
\frac{1}{4!}t_8^{a_1\dots a_8}(\psi_{a_1a_2}\g_i\partial_j\psi_{a_3a_4})
(\psi_{a_5a_6}\g^i\partial^j\psi_{a_7a_8})
-(\psi_{mn}\g^{(n'}\partial^{p')}\psi_{pq})(\psi_{n'}{}^i\g^{[mnp}\partial^{q]}\psi_{p'i})\nn\\
&-(\psi_{n}{}^i\g^{(n'}\partial^{p')}\psi_{pi})(\psi_{n'}{}^j\g^{(n}\partial^{p)}\psi_{p'j})
-\frac14 (\psi_{mn}\g^{[m'n'p'}\partial^{q']}\psi_{pq})(\psi_{m'n'}\g^{[mnp}\partial^{q]}\psi_{p'q'})
\Big\}~.
\end{align}

\subsection*{Pole-subtraction}

All terms in the Lagrangian come with a prefactor $f(s,t,u)$ which  encodes the complete $\al'$ dependence of the 
amplitude. In practice, one is interested in knowing the Lagrangian at 
some given order in the $\al'$ expansion. The first term in the expansion of $f$, as seen in (\ref{f}), has a pole of the form $1/(stu)$. 
This must be subtracted from the Lagrangian, since by unitarity all poles in an $N$-point amplitude must come from 
1-particle-reducible diagrams containing $N^{\prime}$-point vertices, 
where $N^{\prime}<N$. 
However, 
the part proportional to $1/(stu)$ may also contain finite terms which 
contribute to the 4-point Lagrangian. In the NS sector there is a simple derivative-counting 
argument to show that there are no finite parts \cite{grosssloan}. 
 On the other hand, in the RR sector this can no longer be the case, 
as we expect to find 4-point couplings coming 
from the shifts (\ref{modified}) in the 3-point Lagrangian.

Let us illustrate the above discussion using as a concrete example 
the $C_{(0)}^{2} |H^{2}|$ term. 
The counting of derivatives shows that this coupling must come from the pole 
in $f$, since the finite piece contributes an amplitude $\sim F^{2} H^{2}$. 
The relevant part of the amplitude in (\ref{pipb}), namely  $H^{2} F_{(1)}^{2}$,  yields: 
\beal\label{C2H2}
- 2 \sqrt{2}\kappa \, \frac{t^{2}+u^{2}}{s t u} k_{3}^{i} k_{3}^{j} \, (H_{1})_{ilm} 
(H_{2})_{j}^{~lm}  C^{0}_{3} C^{0}_{4} \,.
\end{align}
In order to arrive at this simple form, one must make use of the following relations which
 hold on-shell by virtue of the Bianchi identities:
\beal
k_{3}^{i} k_{4}^{j} (H_{1})_{ilm} (H_{2})_{j}^{~lm} = & (\frac{u}{6} \delta^{ij} - k_{3}^{i} k_{3}^{j})  (H_{1})_{ilm} (H_{2})_{j}^{~lm}\, \nn\\
 (k_{3}^{i} k_{4}^{j} - k_{3}^{j} k_{4}^{i}) (H_{1})_{ilm} (H_{2})_{j}^{~lm} = & (\frac{u-t}{6}) H_{1}\cdot H_{2} \,, \nn\\
 k_{1}^{j} k_{2}^{i} (H_{1})_{ilm} (H_{2})_{j}^{~lm} = & - \frac{s}{6} H_{1}\cdot H_{2}  \,. 
 \end{align}

There are three 1-particle-reducible diagrams which are relevant to this amplitude:

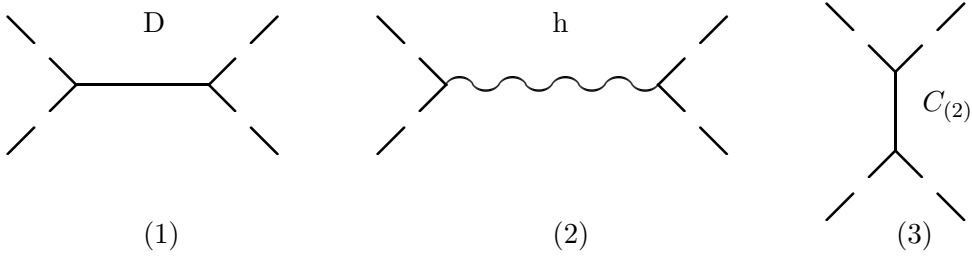
\begin{picture}(10000,14000)
\THICKLINES
\drawline\fermion[\E\REG](6000,8000)[5000]
\put(\pmidx,10000){D}
\put(\pmidx,2000){(1)}
\drawline\scalar[\NW\REG](\fermionfrontx,\fermionfronty)[2]
\drawline\scalar[\SW\REG](\fermionfrontx,\fermionfronty)[2]
\drawline\scalar[\NE\REG](\fermionbackx,\fermionbacky)[2]
\drawline\scalar[\SE\REG](\fermionbackx,\fermionbacky)[2]
\drawline\photon[\E\REG](20000,8000)[8]
\put(\pmidx,10000){h}
\put(\pmidx,2000){(2)}
\drawline\scalar[\NW\REG](\photonfrontx,\photonfronty)[2]
\drawline\scalar[\SW\REG](\photonfrontx,\photonfronty)[2]
\drawline\scalar[\NE\REG](\photonbackx,\photonbacky)[2]
\drawline\scalar[\SE\REG](\photonbackx,\photonbacky)[2]
\drawline\fermion[\N\REG](37000,5500)[3000]
\put(38000,\pmidy){$C_{(2)}$}
\put(37000,2000){(3)}
\drawline\scalar[\SW\REG](\fermionfrontx,\fermionfronty)[2]
\drawline\scalar[\SE\REG](\fermionfrontx,\fermionfronty)[2]
\drawline\scalar[\NW\REG](\fermionbackx,\fermionbacky)[2]
\drawline\scalar[\NE\REG](\fermionbackx,\fermionbacky)[2]
\end{picture}

These diagrams contribute respectively:
\beal
F_{1} = & c_{1} \, \delta^{ij} \,, \nn\\
F_{2} = & c_{2} \, \frac{t u}{s t u} k_{3}^{i} k_{3}^{j} \,, \nn\\
F_{3} = & c_{3} \, \big(\frac13 \delta^{ij} + \frac{s^{2}}{s t u} k_{3}^{i} k_{3}^{j} \big) \,. \nn
\end{align}
One can then check that the sum of the diagrams reproduces the amplitude (\ref{C2H2}) 
exactly up to the contact term generated by $| F_{(3)} - 2 \kappa C_{(0)} H |^{2}$. 
The upshot is that the terms multiplying the 
singular part in the expansion of $f$ are either generated by 1-particle-reducible diagrams, 
or are accounted for by the modification of the RR field strength. 
Only the regular part of $f$ enters the quartic part of the Lagrangian, namely, 
\beal
\widehat{\mathcal{G}}(s,t,u) := f (s,t,u)  + \frac{8 \pi}{{\al'}^{3} stu} \,.
\label{gargl}
\end{align}

%
%\begin{align}
%2 A_{1} + A_{2} + \frac12 A_{3}  & \rightarrow - \sqrt{2} \kappa^{2} (t^{2}+u^{2}) k_{3}^{i}k_{3}^{j} (H_{1})_{i} (H_{2})_{j} C_{3} C_{4}\nn\\
%t_{8}t_{8} R_{1} \ldots R_{4} & \rightarrow 32 \kappa^{4} (t^{2}+u^{2}) k_{3}^{i}k_{3}^{j} (H_{1})_{i} (H_{2})_{j} D_{3} D_{4} 
%\end{align}

\vfill\break

\subsection*{The Lagrangian}

Collecting all the previous subsectors, taking the expansions (\ref{fexp}) into account, 
we finally arrive at the complete four-point Lagrangian:
\beal
{\cal L}_4&=
\frac{1}{2\kappa^2}R-\frac{1}{2}\partial_mD\partial^mD
-\frac{1}{3!}{e^{-\sqrt{2} \kappa D}}H_{mnp}H^{mnp} 
-\frac{1}{\sqrt{2}\kappa }\sum_M\frac{1} {M!}e^{\frac{5-M}{\sqrt{2}} \kappa D}  
\widehat{F}_{m_1\dots m_M}\widehat{F}^{m_1\dots m_M}   \nn\\
&~~~~~~~~~~~~~~~~~~~~~~~~~~~~~~~~~~~~~~~~~~~~~~~~~~~~~~~~~~
+\sqrt{2}\sum_{M+N=8}(-1)^{[\frac{M+1}{2}]}\star(B\wedge F_{(M)}\wedge F_{(N)})+{\cal O}(\psi^2)\nn\\
& +~~\widehat{\mathcal{G}}(\partial, \alpha^{\prime}) \Big\{\frac{1}{4!}
t^{a_1\dots a_8}t_{b_1\dots b_8}\widehat{R}_{a_1a_2}{}^{b_1b_2}\widehat{R}_{a_3a_4}{}^{b_3b_4}
\widehat{R}_{a_5a_6}{}^{b_5b_6}\widehat{R}_{a_7a_8}{}^{b_7b_8}\nn\\
&\qquad\qquad +\sum_{M,N}u_{ij}{}^{mn pq  m'n'p'q'; a_1\dots a_M;b_1\dots b_N}
\widehat{R}_{mnm'n'}
\widehat{R}_{pqp'q'} 
\partial^i{F}_{a_1\dots a_M}\partial^j{F}_{b_1\dots b_N}\nn\\
&\qquad\qquad+\sum_{M,N,P,Q}v^{a_1\dots a_M;b_1\dots b_N;c_1\dots c_P;d_1\dots d_Q}
\partial_{i}\partial_{j}{F}_{a_1\dots a_M}\partial^{i}\partial^{j}{F}_{b_1\dots b_N}
{F}_{c_1\dots c_P}{F}_{d_1\dots d_Q}\nn\\
&\qquad\qquad - 64 i \sum_{M,N}\frac{c_Mc_N\varepsilon_N}{M!N!}\partial^j
F_{a_1\dots a_M}\partial^{(n|}F_{b_1\dots b_N}(\psi_{nk}\g^i\g^{a_1\dots a_M}  \g^{|p)}
\g^{b_1\dots b_N} 
\g_i\partial_j\psi^k{}_p)\nn\\
&\qquad\qquad -32  i \sum_{M,N}\frac{c_Mc_N\varepsilon_N}{M!N!}\partial^j
F_{a_1\dots a_M}\partial^{[m|}F_{b_1\dots b_N}
(\psi_{mn}\g^i\g^{a_1\dots a_M}  \g^{|npq]}\g^{b_1\dots b_N}
\g_i\partial_j\psi_{pq})\nn\\
&\qquad\qquad + 2 it_8^{a_1\dots a_8}R_{a_1a_2n}{}^iR_{a_3a_4pi} (\psi_{a_5a_6}\g^{n}\partial^{p}\psi_{a_7a_8})
+ it_8^{a_1\dots a_8}R_{a_1a_2mn}R_{a_3a_4pq} (\psi_{a_5a_6}\g^{mnp}\partial^{q}\psi_{a_7a_8})\nn\\
&\qquad\qquad +\frac{1}{3}t_8^{a_1\dots a_8}(\psi_{a_1a_2}\g_i\partial_j\psi_{a_3a_4})(\psi_{a_5a_6}\g^i\partial^j\psi_{a_7a_8})
-8(\bar \psi_{mn}\g^{(n'}\partial^{p')}\bar \psi_{pq})(\psi_{n'}{}^i\g^{[mnp}\partial^{q]}\psi_{p'i})\nn\\
&\qquad\qquad 
-8(\bar \psi_{n}{}^i\g^{(n'}\partial^{p')}\bar \psi_{pi})(\psi_{n'}{}^j\g^{(n}\partial^{p)}\psi_{p'j})
-2(\bar \psi_{mn}\g^{[m'n'p'}\partial^{q']}\bar \psi_{pq})(\psi_{m'n'}\g^{[mnp}\partial^{q]}\psi_{p'q'})\nn\\
&\qquad\qquad +(\psi\longleftrightarrow \bar{\psi})\Big\}
~,
\label{thelagrangian}
\end{align}
where the tensors $u$, $v$, are defined in appendix \ref{apptraces}. 
The sums over $M,\dots Q$, run over 
even integers from zero to four for IIA supergravity, 
and over odd integers from one to five for IIB. The action of the operator  
$\widehat{\mathcal{G}}$, cf. (\ref{gargl}), should be understood in the same way as the action of 
${\mathcal{G}}$ in (\ref{drept}).

\section{The linearized superfield}\label{lit}

For type IIB at order $(\alpha')^3$ (i.e. eight
derivatives, or, $R^4$) we can compare our result to the prediction of the
`linearized superfield' of \cite{hw}: 
at the linearized level, in ten dimensional IIB superspace, 
one can define the analogue of a chiral scalar superfield. This 
is sometimes called the linearized, or the analytic, scalar superfield. 
We shall denote it here by ${A}$; it obeys the constraints
\beal
\bar{D}_\al A&=0\nn\\
D^4{A}&=\bar{D}^4\bar{{A}}~.
\end{align}
These constraints restrict the $\theta$-expansion of $A$ to the physical fields of IIB 
supergravity. In particular, the $\theta=0$ component is a complex scalar, the 
$\theta^2$ component is a complex three-form, while the $\theta^4$ component 
contains both the Riemann tensor and $\partial F_{(5)}$, and no new fields appear 
at higher orders in the $\theta$-expansion.  
If ${A}$ is the fluctuation around the flat-space solution, we
expect the action
\beal
\int d^{10}x \int d^{16}\theta {A}^4~
\label{linsup}
\end{align}
to capture the four-field part of the action of type IIB at order $(\alpha')^3$, at the
linearized level. Indeed, note that the action above is supersymmetric, 
up to terms quintic in the fields, and up to terms that 
vanish by virtue of the lowest-order (in $\alpha'$) equations of motion. 
Moreover, note that the $\theta$-integration results in 
eight derivatives (on the bosonic part of the action).

In \cite{gg} it was argued, taking into account the constraints coming from the 
$SL(2,\mathbb{Z})$ invariance of type IIB, that the 
complete, to all string loops, action at order $(\alpha')^3$ is of the form 
\beal
S_{(\alpha')^3}=f(\tau, \bar{\tau})\Big\{t_8t_8R^4+\dots \Big\}~,
\end{align}
where $\tau$ is the axiodilaton, and the ellipses stand for the remaining terms 
in the superinvariant.  
Unfortunately, it is not possible to use the linearized superfield to go beyond the
four-field approximation\footnote{
A chiral measure in on-shell IIB superspace 
does not exist \cite{bh, skenderis}.}. 
For example, the action
\beal
\int d^{10}x \int d^{16}\theta {A}^5~,
\end{align}
would mix with (\ref{linsup}), 
due to the nonlinear terms in the $\theta$-expansion of (the
full-fledged, non-linear extension of) ${A}$. Such terms are 
explicitly discussed in \cite{skenderis}. In going to 
quintic, or higher, order of interactions these nonlinear 
effects can no longer be ignored. This was explicitly verified by the 
authors of \cite{chiral}, who observed that the linearized superfield 
cannot reproduce the $R^2H^3$ terms in the string-theory effective action.

The authors of \cite{pw} observed that their result for the $(\partial F_{(5)})^2R^2$ terms 
in the string-theory effective action, is compatible with the prediction of (\ref{linsup}). 
Let us review the argument: it was found in \cite{pw} that the $(\partial F_{(5)})^2R^2$ terms 
coming from string theory, can be written (by an appropriate field redefinition which 
amounts to setting $\lambda=16$ in formula (2.13) of that reference)  
in such a way that only the $(00200)$, $(20011)$, $(40000)$ representations 
occur\footnote{We are using the Dynkin notation for the 
complexification $D_5$ of $SO(1,9)$. Hence, $(00000)$ is a scalar, 
$(10000)$ is a vector, $(01000)$ is a two-form, etc.} 
in the 
tensor-product decomposition of $R^2$. On the other hand, from the point-of-view 
of the linearized superfield, these terms come from  (taking 
into account the fact 
that Grassmann integration can be thought of as differentiation)
$$(D^4A)^2(D^4A)^2\sim (\partial F_{(5)})^2 R^2~.$$ 
It follows that only representations in $(00001)^{8\otimes_a}\cap (02000)^{2\otimes_s}$ can 
occur in the decomposition of $R^2$. However, one can see that 
$$ (00001)^{8\otimes_a}\cap (02000)^{2\otimes_s} =(00200)\oplus(20011)\oplus(40000)~, $$
in agreement with string theory.

Note that a similar representation-theoretic argument 
cannot be used to compare the $(\partial F_{(1)})^2R^2$, $(\partial F_{(3)})^2R^2$ 
terms in the string-theory effective action to the corresponding terms in the linear-superfield 
action. The reason is that the two-axion, two-graviton terms in (\ref{linsup}) come from 
$$A(D^8A)(D^4A)^2\sim C_{(0)}\partial^4C_{(0)}R^2~,$$ 
and can only be compared to the string-theory result 
after partial integration. Similarly the two-threeform, two-graviton terms 
in (\ref{linsup}) come from 
$$(D^2A)(D^6A)(D^4A)^2\sim F_{(3)}\partial^2F_{(3)}R^2~.$$

\section{Discussion}\label{discussion}

It would be of interest to try to lift the ten-dimensional type IIA 
$R^4$-correction to eleven dimensions\footnote{Spinorial-cohomology techniques \cite{cnt} 
can be employed to show that the first higher-order correction in eleven dimensions 
appears at order $l_{Planck}^3$ \cite{lthree} (five derivatives). 
This correction is of topological 
nature and is related to the shifted quantization condition 
of the four-form field strength in eleven dimensions \cite{witten}.}. Partial results on such an 
attempt were reported  in \cite{pvwreview} and more 
recently in \cite{gva, gvb} (see also \cite{mafra} which addresses 
some problems with the computation of \cite{gva}). 
At present, in the absence 
of a superPoincar\'{e}-invariant microscopic formulation of M-theory, 
a covariant computation directly in eleven dimensions 
would have to rely on supersymmetry\footnote{ 
Some results on this have appeared recently in 
\cite{roufatokavli}. These authors use the N\"{o}ther method to 
partially cancel the supersymmetric variation 
of a certain subsector of the action at the eight-derivative order.}. 
The current status of the superspace 
approach to higher-order derivative corrections in eleven dimensions 
can be found in \cite{cgnt}, in which the supertorsion Bianchi identities in eleven 
dimensions are solved in all generality to first order 
in a deformation parameter (see \cite{cgnn} for earlier partial results). 
In \cite{cgnt} the deformations to the 
supertorsion constraints were parameterized in terms of 
certain superfields which were treated as `black boxes'. In order 
to obtain explicit expressions however, these
superfields would ultimately have to be 
expressed in terms of the physical fields in the massless multiplet. 
Unfortunately at present this remains a very difficult problem,  
equivalent to the computation of certain spinorial-cohomology groups, 
although a systematic way to arrive at these explicit corrections 
has been proposed in \cite{ht}.

An obvious extension of the results in this paper is the investigation of higher-point, 
eight-derivative couplings. 
With the exception of the anomaly-related 
Chern-Simons terms in ten or eleven dimensions \cite{md} (which appear at five points), 
this is a subject about which very little is known 
(see \cite{chiral} for some partial results, and \cite{pw} for a general discussion).  
It would be of interest to examine whether the factorization property 
(\ref{tensor}) can be generalized in any useful way  
to the case of quintic, or higher, interactions.

An important application of higher-order corrections, one which has recently attracted 
renewed interest, is the modification of the macroscopic properties of black-holes. 
String theory provides an underlying  microscopic formulation 
within which the thermodynamic properties of black holes (at least of certain kinds thereof) 
can be derived with remarkable accuracy. It has been observed that  
higher-derivative ($R^2$ in four spacetime dimensions) 
contributions may lead to qualitatively different behavior, for example 
the appearance of a horizon even in the case where some of the 
black-hole charges vanish. It would be of interest to investigate 
the implications  for this subject, 
of the higher-derivative corrections derived in the present paper.

In principle our result can be used to compute the 
corrections to the supersymmetry transformations to all orders in $\alpha^{\prime}$, 
at the quartic approximation in the fields. This is of interest to the investigation 
of higher-derivative corrections to supersymmetric backgrounds. In particular, it would be 
desirable to discover contexts in which  these corrections can qualitatively modify the 
geometrical properties of the background, e.g. smooth-out singularities, etc.  
We hope to report on this in the future.

\vfill\break

\section*{Acknowledgments} 

We are grateful to Nathan Berkovits, Paul Howe and Kasper Peeters for useful email 
correspondence. D.T. would like to 
thank the organizers of the Workshop on Pure-Spinor Formalism in String Theory, 
IHES, Bures-sur-Yvette, France, for a stimulating atmosphere.

\appendix

\section{Fierz identities}\label{fierz}

For anticommuting spinors $\q^\al,\phi^\be$, 
\beal
\q^\al\phi^\be =\frac1{16} (\g_m)^{\al\be} (\q \g^m \phi) +
\frac{1}{96}(\g_{mnp})^{\al\be}(\q\g^{mnp}\phi)~+ \frac1{3840}
(\g_{mnpqr})^{\al\be} (\q\g^{mnpqr}\phi) ,  
\end{align}
from which it follows that
\beal
(\q M\chi)(\q N\psi)=-\frac{1}{96}(\q \g^{mnp}\q)(\chi M^{tr}\g_{mnp}N\psi)~.
\end{align}
Using the above we can prove the following Fierz identities
\beal
(\g^{[i}{}_{mn}\q)_\al(\q\g^{j]mn}\q)&=2(\g_k\q)_\al(\q\g^{ijk}\q)\nn\\
(\g^{ijmnp}\q)_\al(\q\g_{mnp}\q)&=-6(\g_k\q)_\al(\q\g^{ijk}\q)
~.
\label{bakunin}
\end{align}

\section{Zero-mode formul\ae}\label{zeromode}

The following is list a of formul\ae{} repeatedly used in 
the derivation of the amplitudes:

\beal
T^{\al\be\g}
(\g^i\q)_\al(\g^j\q)_\be(\g^k\q)_\g(\q \g^{i'j'k'}\q) F_{ijk,i'j'k'}=
\frac{1}{120} F_{ijk,}{}^{ijk}~,
\label{kropotkin}
\end{align}
where $F_{ijk,i'j'k'}$ is any tensor 
antisymmetric in the first three and in the last three indices and 
\beal
T^{\al\be\g}:=
T_{\rho\sigma\tau}^{\al\be\g}
(\frac{\del}{\del\q}\g^{pmn}\frac{\del}{\del\q})
(\g_p\frac{\del}{\del\q})^\rho
(\g_m\frac{\del}{\del\q})^\sigma
(\g_n\frac{\del}{\del\q})^\tau 
~.
\end{align}
\beal
T^{\al\be\g}
(\g^{imn}\q)_\al(\g^j\q)_\be(\g^k\q)_\g(\q \g^{i'j'k'}\q) F_{imn,jk,i'j'k'}
=\frac{1}{70} F_{imn,}{}^i{}_{j,}{}^{jmn}~,
\label{luxemburg}
\end{align}
where $F_{imn,jk,i'j'k'}$ is any tensor 
antisymmetric in $(i,m,n)$, in $(j,k)$ and $(i',j',k')$.%
%\beal
%T^{\al\be\g}
%(\g^i \g^m \g^n\q)_\al(\g^j\q)_\be(\g^k\q)_\g(\q \g^{i'j'k'}\q)
%F_{i,m,n,jk,i'j'k'} = \nn\\ 
%\frac{1}{70} F_{[i,m,n],}{}^i{}_{j,}{}^{jmn} + \frac1{120}
%(F_{i,m,m,j,k,i,j,k} +F_{m,m,i,j,k,i,j,k} - F_{i,m,i,j,k,m,j,k}) \, .
%\label{luxemburg2}
%\end{align}
%
\beal
T^{\al\be\g}
(\g^{m_1\dots m_5}\q)_\al(\g^j\q)_\be&(\g^k\q)_\g(\q \g^{i'j'k'}\q)
F_{m_1\dots m_5,jk,i'j'k'}\nn\\ 
&=-\frac{1}{42}\Big\{ F^{m_1\dots m_5}{}_{,m_1m_2,m_3m_4m_5} +\frac{1}{5!}
\varepsilon^{m_1\dots m_{10}}F_{m_1\dots m_5,m_6m_7,m_8m_9m_{10}}\Big\}~,
\label{trotsky}
\end{align}
where $F_{m_1\dots m_5,jk,i'j'k'}$ is any tensor 
antisymmetric in $(m_1\dots m_5)$, in $(j,k)$ and $(i',j',k')$. 
Similarly, for an antichiral spinor $\bar{\theta}_\al$ we have:
\beal
T_{\al\be\g}
(\g^{m_1\dots m_5}\bar{\q})^\al(\g^j\bar{\q})^\be
&(\g^k\bar{\q})^\g(\bar{\q}\g^{i'j'k'}\bar{\q})  
F_{m_1\dots m_5,jk,i'j'k'}\nn\\
&=-\frac{1}{42}\Big\{ F^{m_1\dots m_5}{}_{,m_1m_2,m_3m_4m_5} -\frac{1}{5!}
\varepsilon^{m_1\dots m_{10}}F_{m_1\dots m_5,m_6m_7,m_8m_9m_{10}}\Big\}~.
\label{trotskyb}
\end{align}
We can prove (\ref{kropotkin}) as follows. First note that the left-hand side 
is a scalar. On the other hand there is only one scalar in the tensor
product of  
two three-forms
$$
F_{ijk,i'j'k'}\sim(00100)^{2\otimes}= 1(00000)\oplus\dots
$$
and we can take the right-hand side to be proportional to 
$F_{ijk,}{}^{ijk}$. The proportionality constant is determined by taking 
$ F_{ijk,}{}^{i'j'k'}=\delta^{i'}_{[i}\delta^{j'}_{j}\delta^{k'}_{k]}$ 
and noting that $\delta^{i}_{[i}\delta^{j}_{j}\delta^{k}_{k]}=120$.

Similarly, we can prove (\ref{luxemburg}) by noting that there is only one 
scalar in the decomposition of the tensor product of two three-forms
and a two-form: 
$$
F_{imn,jk,i'j'k'}\sim(00100)^{2\otimes}\otimes(01000)= 1(00000)\oplus\dots
$$
and we can take the right-hand side to be proportional to 
$F_{imn,}{}^i{}_{j,}{}^{jmn}$. In order to determine 
the proportionality constant we set  
$$
F_{imn,}{}^j{}_{k,}{}^{i'j'k'}=\frac{1}{2}\Big(
\delta^{[i'}_{[m}\delta^{j'}_{n|}\delta^{k']}_k\delta^j_{|i]}
-\delta^{[i'}_{[m}\delta^{j'}_{n|}\eta^{k']j}\eta_{|i]k}
\Big)~,
$$
so that
$$
T^{\al\be\g}
(\g^{imn}\q)_\al(\g^j\q)_\be(\g^k\q)_\g(\q \g^{i'j'k'}\q) F_{imn,jk,i'j'k'}
=T^{\al\be\g}
(\g^{jmn}\q)_\al(\g^j\q)_\be(\g^k\q)_\g(\q \g^{kmn}\q) ~.
$$
We then arrive at (\ref{luxemburg}) by taking (\ref{bakunin}) into
account and noting that  
$$
\frac{1}{2}\Big(
\delta^{[m}_{[m}\delta^{n}_{n|}\delta^{k]}_k\delta^j_{|j]}
-\delta^{[m}_{[m}\delta^{n}_{n|}\eta^{k]j}\eta_{|j]k}
\Big)=140~.
$$
A consequence of (\ref{kropotkin}, \ref{luxemburg}) is the following formula
\beal
T^{\al\be\g}
(\g^{i}\g^{mn}\q)_\al(\g^j\q)_\be(\g^k\q)_\g&(\q \g^{i'j'k'}\q)
F_{i,mn,jk,i'j'k'}\nn\\ 
& = \frac{1}{210} F_{i,mn;}{}^i{}_k,{}^{kmn} - \frac{1}{105}
F_{m,in,}{}^i{}_k,{}^{kmn}    
+\frac{1}{60}F^m{}_{,mi,jk,}{}^{ijk}~,
\label{uljanof}
\end{align}
where $F_{i,mn,jk,i'j'k'}$ is any tensor 
antisymmetric in $(m,n)$, in $(j,k)$ and $(i',j',k')$.

We can prove (\ref{trotsky})  by noting that there are two  
scalars in the decomposition of the tensor product of a five-form, a three-form and a two-form. 
Hence we can take the right-hand side to be proportional to 
$$\alpha F^{m_1\dots m_5}{}_{,m_1m_2,m_3m_4m_5} +\beta 
\varepsilon^{m_1\dots m_{10}}F_{m_1\dots m_5,m_6m_7,m_8m_9m_{10}}~.$$ 
In order to determine 
the constant $\alpha$ we set $ F^{m_1\dots m_5}{}_{,m_1m_2,m_3m_4m_5}= 
\delta_{[m_6}^{m_1}\dots\delta_{m_{10}]}^{m_5}$, taking (\ref{bakunin}) into account 
and noting that 
$$
\delta_{[m_1}^{m_1}\dots\delta_{m_{5}]}^{m_5}=252~. 
$$
To determine $\beta$ we set  
$F_{m_1\dots m_{10}}= \varepsilon_{m_1\dots m_{10}}$, taking the Hodge dualization 
\beal
\g_{(n)}=(-)^{\frac{1}{2}n(n-1)}*\g_{(10-n)}\g_{11}
\label{mayakovsky} 
\end{align}
into account. Equation (\ref{trotskyb}) is proven similarly.

\section{Amplitudes}\label{amplitudes}

We can break the correlator (\ref{A4}) down according to the individual terms in
$V_4$, as follows: 
 
$\bullet$ $\del\q^\al A_\al$: This does not contribute.

$\bullet$ $\Pi^m A_m$: In this correlator, $\Pi^m \sim \del x^m$. 
We can separate the contributions according to the number of $\theta$'s 
in the vertices. 

\bit 
\item $\beta_1$ : $A_\al^{(1)} A_\beta^{(1)}A_\g^{(1)}A_m^{(2)}$ 

\beal 
%F = & - \frac{\al'}{32} \Pi(z_{ij}) a_{[i}(1) a_j(2) a_{k]}(3) \left(
%\sum_{r}^3 \frac{k_{i'}(r)}{z_r - z_4} \right) a_{j'}(4) k_{k'}(4) \,;
%\nn \\
 - \frac{\al'}{23040} \Pi(z_{ij}) &\left[ (\frac{a_1 \cdot 
    k_2}{z_2-z_4} + \frac{a_1 \cdot k_3}{z_3 - z_4}) (a_2 \cdot a_4 \,
  a_3 \cdot k_4 - a_2 \cdot k_4 \, a_3 \cdot a_4) + \right. \nn \\ 
& \left. - \, (\frac{a_2 \cdot 
    k_1}{z_1-z_4} + \frac{a_2 \cdot k_3}{z_3 - z_4}) (a_1 \cdot a_4 \,
  a_3 \cdot k_4 - a_1 \cdot k_4 \, a_3 \cdot a_4) + \right. \nn \\
& \left. + (\frac{a_3 \cdot 
    k_1}{z_1-z_4} + \frac{a_3 \cdot k_2}{z_2 - z_4}) (a_1 \cdot a_4 \,
  a_2 \cdot k_4 - a_1 \cdot k_4 \, a_2 \cdot a_4) \right] \, .
\end{align}

\item $\beta_2$ : $A_\al^{(2)} A_\beta^{(1)}A_\g^{(1)}A_m^{(1)}$

\beal 
 - \frac{i \al'}{34560} \frac{\Pi(z_{ij})}{z_1-z_4} & \Big\{
- \xi_2 [\slsh k_1, \slsh a_1 ] \slsh a_3 \xi_4 +2a_1\cdot a_3~\xi_2
\slsh k_1\xi_4  
+ \xi_3 [\slsh k_1 , \slsh a_1 ] \slsh a_2 \xi_4 \nn\\ 
&-2a_1\cdot a_2~\xi_3 \slsh k_1\xi_4 
+4  k_1 \cdot
  a_2 (\xi_3 \slsh a_1 \xi_4 - \xi_1 \slsh a_3 \xi_4) \nn\\ 
&+ 4 k_1 \cdot a_3
  (\xi_1 \slsh a_2 \xi_4 
- \xi_2 \slsh a_1 \xi_4)  \Big\} - (2,1,3) - (3,2,1) ~.
\end{align}

\item $\beta_3$ : $A_\al^{(3)} A_\beta^{(1)}A_\g^{(1)}A_m^{(0)}$

\beal 
 - \frac{\al'}{5760} \Pi(z_{ij}) \sum_r^3 \frac{k_r
  \cdot a_4}{z_r-z_4} 
[ a_1 \cdot a_2 &\, (k_1-k_2) \cdot a_3 \nn\\
&+ a_1 \cdot a_3 \, (k_3 -  k_1) \cdot a_2 + a_2 \cdot a_3 \, (k_2 -
k_3) \cdot a_1 ]  
\end{align}

\item $\beta_4$ : $A_\al^{(2)} A_\beta^{(2)}A_\g^{(1)}A_m^{(0)}$ 
\beal
  \frac{i \al'}{2880} \Pi(z_{ij}) \sum_r^3 \frac{k_r \cdot
  a_4}{z_r-z_4} \left(\xi_1 \slsh a_3 \xi_2 - \xi_1 \slsh a_2 \xi_3
  +\xi_2 \slsh a_1 \xi_3 \right) \,.  
\end{align} 

\eit 
 
$\bullet$ $d_\al W^\al$ : here $d_\al$ contributes with $p_\al$ and
 $\del x^m (\theta \g_m)_\al$. The latter is easy to compute 
observing that, from the $\theta$ expansion, $\theta \g_m
 W^{(0)} = A_m^{(1)}$, and $\theta \g_m W^{(1)} = 2
 A_m^{(2)}$. Therefore these terms give $- \beta_1 - \frac12
 \beta_2$. 

The terms with $p$ are: 

\bit

\item $\g_1$ : $U^{(1)}  U^{(1)}  U^{(1)} W^{(3)}$

The correlator is equal to:
\beal
-\frac{\al' }{24\times 640}\Pi(z_{ij})&\Big\{
\frac{a_1\cdot k_4}{z_1-z_4}[a_2\cdot k_4~a_3\cdot a_4-a_3\cdot
  k_4~a_2\cdot a_4 ]\nn\\ 
-&\frac{a_2\cdot k_4}{z_2-z_4}[a_1\cdot k_4~a_3\cdot a_4-a_3\cdot
  k_4~a_1\cdot a_4 ]\nn\\ 
-&\frac{a_3\cdot k_4}{z_3-z_4}[a_2\cdot k_4~a_1\cdot a_4-a_1\cdot
k_4~a_2\cdot a_4 ] 
\Big\}~.
\end{align}

\item $\g_2$ : $U^{(2)}  U^{(1)}  U^{(1)} W^{(2)}$ 

\beal
\frac{i \al'}{144 \times 240}  \Pi(z_{ij})\Big\{ &  
  \frac{1}{z_1-z_4} \, \frac{23}{4}(k_4 \cdot a_2 \, \xi_1 \slsh a_3 \xi_4 -
  k_4 
\cdot a_3 \, \xi_1 \slsh a_2 \xi_4 ) \nn \\ 
- &  \frac1{z_2-z_4} \,  ( 4 k_4 \cdot a_2 \, \xi_1 \slsh a_3
\xi_4  +  k_4 \cdot a_3 \, \xi_1 \slsh a_2 \xi_4 ) \nn \\
+ &  \frac1{z_3-z_4} \, ( k_4 \cdot a_2 \, \xi_1 \slsh a_3
  \xi_4 + 4  k_4 \cdot a_3 \, \xi_1 \slsh a_2 \xi_4 )    
\Big\}  \nn\\
 & - (2,1,3) - (3,2,1) 
\end{align}

\item $\g_3$ : $U^{(3)}  U^{(1)}  U^{(1)} W^{(1)}$ 

The correlator is equal to:
\beal
\frac{\al' }{192\times 240}\Pi(z_{ij})\Big\{
\frac{17}{2}&\frac{1}{z_1-z_4}[-a_1\cdot k_4~a_2\cdot k_1~a_3\cdot a_4
+a_1\cdot k_4~a_3\cdot  k_1~a_2\cdot a_4 \nn\\ 
&\qquad\quad-a_2\cdot k_4~a_3\cdot k_1~a_1\cdot a_4
+a_2\cdot k_4~a_4\cdot k_1~a_1\cdot a_3\nn\\
&\qquad\quad-a_3\cdot k_4~a_4\cdot k_1~a_1\cdot a_2
+a_3\cdot k_4~a_2\cdot k_1~a_1\cdot a_4\nn\\
&\qquad\quad-k_1\cdot k_4~a_1\cdot a_3~a_2\cdot a_4
+k_1\cdot k_4~a_1\cdot a_2~a_3\cdot a_4
]\nn\\
+&\frac{1}{z_2-z_4}[-a_1\cdot k_4~a_2\cdot k_1~a_3\cdot a_4
-4 a_1\cdot k_4~a_3\cdot  k_1~a_2\cdot a_4 \nn\\ 
&\qquad\quad+a_1\cdot k_4~a_4\cdot k_1~a_2\cdot a_3
+4a_2\cdot k_4~a_3\cdot k_1~a_1\cdot a_4\nn\\
&\qquad\quad-4a_2\cdot k_4~a_4\cdot k_1~a_1\cdot a_3
- a_3\cdot k_4~a_4\cdot  k_1~a_1\cdot a_2 \nn\\ 
&\qquad\quad+a_2\cdot k_1~a_3\cdot k_4~a_1\cdot a_4
-k_1\cdot k_4~a_1\cdot a_4~a_2\cdot a_3\nn\\
&\qquad\quad+4k_1\cdot k_4~a_1\cdot a_3~a_2\cdot a_4
+k_1\cdot k_4~a_1\cdot a_2~a_3\cdot a_4
]\nn\\
-&\frac{1}{z_3-z_4}[-a_1\cdot k_4~a_3\cdot k_1~a_2\cdot a_4
-4 a_1\cdot k_4~a_2\cdot  k_1~a_3\cdot a_4 \nn\\ 
&\qquad\quad+a_1\cdot k_4~a_4\cdot k_1~a_2\cdot a_3
+4a_3\cdot k_4~a_2\cdot k_1~a_1\cdot a_4\nn\\
&\qquad\quad-4a_3\cdot k_4~a_4\cdot k_1~a_1\cdot a_2
- a_2\cdot k_4~a_4\cdot  k_1~a_1\cdot a_3 \nn\\ 
&\qquad\quad+a_3\cdot k_1~a_2\cdot k_4~a_1\cdot a_4
-k_1\cdot k_4~a_1\cdot a_4~a_2\cdot a_3\nn\\
&\qquad\quad+4k_1\cdot k_4~a_1\cdot a_2~a_3\cdot a_4
+k_1\cdot k_4~a_1\cdot a_3~a_2\cdot a_4]
\Big\}\nn\\
 & - (2,1,3) - (3,2,1) ~.
\end{align}

\item $\g_4$ : $U^{(2)}  U^{(2)}  U^{(1)} W^{(1)}$

\beal 
  \frac{i \al'}{128 \times 27 \times 15} \Pi(z_{ij}) \Big\{&\frac{1}{z_1-z_4}
\Big[ 2 a_3\cdot k_4~\xi_1 \slsh a_4 \xi_2 -2 a_3\cdot a_4~\xi_1 \slsh k_4
  \xi_2 - 4  \xi_1 [\slsh k_4, \slsh a_4
  ] \slsh a_3 \xi_2 \Big]\nn\\
- & \frac{1}{z_2-z_4}
\Big[ 2 a_3\cdot k_4~\xi_2 \slsh a_4 \xi_1 -2 a_3\cdot a_4~\xi_2 \slsh k_4
  \xi_1 - 4  \xi_2 [\slsh k_4, \slsh a_4
  ] \slsh a_3 \xi_1 \Big]\nn\\
+&\frac{9}{z_3-z_4}
\Big[ a_3\cdot a_4~\xi_1 \slsh k_4\xi_2 
-a_3\cdot k_4~\xi_1 \slsh a_4\xi_2\Big] \Big\} - (1,3,2) - (3,2,1) ~.
\end{align}

\item $\g_5$ : $U^{(4)}  U^{(1)}  U^{(1)} W^{(0)}$

\beal 
  \frac{i \al'}{ 144} \Pi(z_{ij}) \Big\{-&\frac1{64}\frac{1}{z_1-z_4} 
  (k_1 \cdot a_3 \, \xi_1 \slsh{a_2} \xi_4 - k_1 \cdot a_2 \, \xi_1
  \slsh{a_3} \xi_4 ) + \, \nn\\ 
+ & \frac1{240}\frac1{z_2-z_4} \, k_1 \cdot a_3 \, \xi_1 \slsh{a_2} \xi_4
- \frac1{240}\frac1{z_3-z_4} \,  k_1 \cdot a_2 \, \xi_1 \slsh{a_3} \xi_4
\Big\} \nn\\
& - (2,1,3) - (3,2,1) \, . 
\end{align}

\item $\g_6$ : $U^{(3)}  U^{(2)}  U^{(1)} W^{(0)}$

\beal 
  \frac{i \al'}{96} \Pi(z_{ij}) \Big\{&\frac{1}{z_1-z_4} 
  (\frac1{72} k_1 \cdot a_3 \xi_2 \slsh{a_1} \xi_4 - \frac1{72} a_1
  \cdot a_3 \xi_2 \slsh{k_1} \xi_4 + \frac1{90} k_1^{a} a_1^{b}
  a_3^{c} \xi_2 \g_{abc} \xi_4 ) + \, \nn\\ 
+ &\frac1{z_2-z_4}(- \frac1{60}  k_1 \cdot a_3 \xi_2 \slsh{a_1} \xi_4
+ \frac1{60} a_1  \cdot a_3 \xi_2 \slsh{k_1} \xi_4 - \frac1{1440}
k_1^{a} a_1^{b}  a_3^{c} \xi_2 \g_{abc} \xi_4 ) 
\Big\} \nn\\
+ &  \frac1{z_3-z_4}(\frac1{360}  k_1 \cdot a_3 \xi_2 \slsh{a_1} \xi_4
- \frac1{360} a_1  \cdot a_3 \xi_2 \slsh{k_1} \xi_4 - \frac1{360}
k_1^{a} a_1^{b}  a_3^{c} \xi_2 \g_{abc} \xi_4 )  
\Big\} \nn \\
+&  {\textrm {cyclic perms.}} 
\end{align}

\item $\g_7$ : $U^{(2)}  U^{(2)}  U^{(2)} W^{(0)}$ 

\beal
-\frac{ \al' }{2880}\Pi(z_{ij})&\Big\{
\frac{1}{z_1-z_4} (\xi_1 \g^a \xi_4) \, (\xi_2 \g_a \xi_3) + \nn\\ 
& - \frac{1}{z_2-z_4} (\xi_2 \g^a \xi_4) \, (\xi_1 \g_a \xi_3) +\nn\\ 
& + \frac{1}{z_3-z_4} (\xi_3 \g^a \xi_4) \, (\xi_1 \g_a \xi_2)
\Big\}~.
\end{align}

\eit

$\bullet$ $N_{mn} F^{nm}$: This case is similar to the $\beta$s, as the
contractions involve only the ghosts. One has $\la^\alpha(x) N_{mn}(y)
\sim \frac{\al'}{2(x-y)} (\la \g_{mn})^\al$. 

\bit
 
\item $\delta_1$ : $A_\al^{(1)} A_\beta^{(1)}A_\g^{(1)}A_m^{(2)}$ 

\beal 
 - \frac{\al'}{64 \times 144} \Pi(z_{ij}) & \Big\{
 \frac1{z_1-z_4} (k_4 \cdot a_1 \, k_4 \cdot a_2 \, a_3 \cdot a_4 - k_4
 \cdot a_1 \, k_4 \cdot a_3 \, a_2 \cdot a_4 ) \Big\} \nn\\
& - (2,1,3) - (3,2,1) ~.
\end{align}

\item $\delta_2$ : $A_\al^{(2)} A_\beta^{(1)}A_\g^{(1)}A_m^{(1)}$

\beal
\frac{i \al'}{34560} \Pi(z_{ij}) & \Big\{
 \frac1{z_1-z_4}\frac1{2} (k_4 \cdot a_2 \, \xi_1 \slsh{a_3} \xi_4 -
 k_4 \cdot a_3 \xi_1 \slsh{a_2} \xi_4 ) + \nn\\ 
& -  \frac{6}{z_2-z_4}k_4 \cdot a_2 \, \xi_1 \slsh{a_3} \xi_4 +
\frac{6}{z_3-z_4} k_4 \cdot a_3 \xi_1 \slsh{a_2} \xi_4 ) \Big\} \,
\nn\\
& - (2,1,3) + (3,1,2) ~. 
\end{align}

\item $\delta_3$ : $A_\al^{(3)} A_\beta^{(1)}A_\g^{(1)}A_m^{(0)}$

\beal 
- \frac{\al'}{128 \times 360} \Pi(z_{ij}) & \Big\{ \frac12
\frac1{z_1-z_4} ( - k_4 \cdot a_1 k_1 \cdot a_2 a_3 \cdot a_4 + k_4
\cdot a_1 k_1 \cdot a_3 a_2 \cdot a_4 \nn\\ 
& - k_4 \cdot a_2 k_1 \cdot a_3 a_1 \cdot a_4 + k_4 \cdot a_2 k_1
\cdot a_4 a_1 \cdot a_3 \nn\\
& - k_4 \cdot a_3 k_1 \cdot a_4 a_1 \cdot a_2 + k_4 \cdot a_3 k_1
\cdot a_2 a_1 \cdot a_4 \nn\\ 
& - k_1 \cdot k_4 a_1 \cdot a_3 a_2 \cdot a_4 + k_1 \cdot k_4 a_1
\cdot a_2 a_3 \cdot a_4 ) \nn\\
+ & \frac1{z_2-z_4} ( - k_4 \cdot a_1 k_1 \cdot a_2 a_3 \cdot a_4 + 4
k_4 \cdot a_1 k_1 \cdot a_3 a_2 \cdot a_4 \nn\\
& + k_4 \cdot a_1 k_1 \cdot a_4 a_2 \cdot a_3 -4 k_4 \cdot a_2 k_1
\cdot a_3 a_1 \cdot a_4 \nn\\
& + 4 k_4 \cdot a_2 k_1 \cdot a_4 a_1 \cdot a_3 - k_4 \cdot a_3 k_1
\cdot a_4 a_1 \cdot a_2 \nn\\
& + k_4 \cdot a_3 k_1 \cdot a_2 a_1 \cdot a_4 - k_4 \cdot k_1 a_2
\cdot a_3 a_1 \cdot a_4 \nn\\ 
& -4 k_4 \cdot k_1 a_2 \cdot a_4 a_1 \cdot a_3 + k_4 \cdot k_1 a_1
\cdot a_2 a_3 \cdot a_4 ) \nn\\
& + \frac1{z_3-z_4} ( -4 k_4 \cdot a_1 k_1 \cdot a_2 a_3 \cdot a_4 +
k_4 \cdot a_1 k_1 \cdot a_3 a_2 \cdot a_4 \nn\\
& - k_4 \cdot a_1 k_1 \cdot a_4 a_2 \cdot a_3 - k_4 \cdot a_2 k_1
\cdot a_3 a_1 \cdot a_4 \nn\\ 
& + k_4 \cdot a_2 k_1 \cdot a_4 a_1 \cdot a_3 -4 k_4 \cdot a_3 k_1
\cdot a_4 a_1 \cdot a_2 \nn\\
& + 4 k_4 \cdot a_3 k_1 \cdot a_2 a_1 \cdot a_4 + k_4 \cdot k_1 a_1
\cdot a_4 a_2 \cdot a_3 \nn\\
& - k_4 \cdot k_1 a_1 \cdot a_3 a_2 \cdot a_4 + 4 k_4 \cdot k_1 a_1
\cdot a_2 a_3 \cdot a_4) \Big\} \\ 
& - (2,1,3) + (3,1,2) ~.
\end{align}

\item $\delta_4$ : $A_\al^{(2)} A_\beta^{(2)}A_\g^{(1)}A_m^{(0)}$ 
\beal
\frac{i \al'}{36 \times 96 \times 15} \Big\{ & \frac1{z_1-z_4} ( k_4^a
a_3^b a_4^c \, \xi_1 \g_{abc} \xi_2 + a_3 \cdot a_4 \, \xi_1
\slsh{k_4} \xi_2 - a_3 \cdot k_4 \, \xi_1 \slsh{a_4} \xi_2 ) + \nn\\
+ & \frac1{z_2-z_4} ( - k_4^a
a_3^b a_4^c \, \xi_1 \g_{abc} \xi_2 + a_3 \cdot a_4 \, \xi_1
\slsh{k_4} \xi_2 - a_3 \cdot k_4 \, \xi_1 \slsh{a_4} \xi_2 ) + \nn\\
+ & \frac1{z_3-z_4} ( 9 a_3 \cdot a_4 \, \xi_1
\slsh{k_4} \xi_2 - 9 a_3 \cdot k_4 \, \xi_1 \slsh{a_4} \xi_2 ) \Big\}
+ \nn\\
& - (1,3,2) + (2,3,1) ~.
\end{align} 

\eit

The full 4-boson amplitude comes from $\beta_{3}, \g_{1},\g_{3},\delta_{1},\delta_{3}$. The result is  

\beal 
\frac{\al'}{5760} \Big[ \frac{\Pi(z_{ij})}{z_{1}-z_{4}} & \Big\{  2 \, k_{1} \cdot a_{4} \, k_{2} \cdot a_{3} \, a_{1} \cdot 	a_{2} - 
	2 \, k_{1} \cdot a_{4} \, k_{3} \cdot a_{2} \, a_{1} \cdot a_{3} + \nn \\
	& +2 k_{3} \cdot a_{4} \, k_{2} \cdot a_{1} \, a_{2} \cdot a_{3} -2  k_{2} \cdot a_{4} \, k_{3} \cdot a_{1} \, a_	{2} \cdot a_{3}+ \nn\\
	& +2 k_{1} \cdot a_{3} \, k_{3} \cdot a_{2} \, a_{1} \cdot a_{4} - 2 k_{1} \cdot a_{2} \, k_{2} \cdot a_{3} \, a_ 	{1} \cdot a_{4} + \nn \\
	& -2 k_{4} \cdot a_{1} \, k_{2} \cdot a_{3} \, a_{2} \cdot a_{4}   + 2 k_{4} \cdot a_{1} \, k_{3} \cdot a_{2} \, 
	a_{3} \cdot a_{4}   + \nn \\
	& + k_{1} \cdot k_{4} \, (a_{1} \cdot a_{2} \, a_{3} \cdot a_{4} - a_{1} \cdot a_{3} \, a_{2} \cdot a_{4}) + 
	(k_{3}-k_{2})\cdot k_{4} \, a_{2} \cdot a_{3} \, a_{1} \cdot a_{4} \Big\} \Big] \nn\\
	& - (2,1,3) - (3,2,1) 
\end{align}

The contribution from $\beta_{2},\g_{2},\delta_{2}, \g_{5}, \g_{6}$ is 

\beal 
\frac{i\al'}{24 \times 5760} \Big[ \frac{\Pi(z_{ij})}{z_{1}-z_{4}} & \Big\{ 
	\xi_{1} \slsh{a_{2}} \xi_{4} \, (2 k_{1} \cdot a_{3} + 49 k_{2} \cdot a_{3} ) + 
	\xi_{1} \slsh{a_{3}} \xi_{4} \, (- 2 k_{1} \cdot a_{2} - 49 k_{3} \cdot a_{2} ) \nn\\
	+ & \xi_{2} \slsh{a_{1}} \xi_{4} \, (24 k_{1} \cdot a_{3} -8  k_{2} \cdot a_{3} ) + 
	\xi_{3} \slsh{a_{1}} \xi_{4} \, (-24  k_{1} \cdot a_{2} + 8 k_{3} \cdot a_{2} ) \nn \\
	+ & \xi_{2} \slsh{a_{3}} \xi_{4} \, (-40  k_{2} \cdot a_{1} -44 k_{3} \cdot a_{1} ) +
	\xi_{3} \slsh{a_{2}} \xi_{4} \, (44 k_{2} \cdot a_{1} + 40 k_{3} \cdot a_{1} ) \nn\\
	- & 24 a_{1} \cdot a_{3} \xi_{2} \slsh{k_{1}} \xi_{4} + 24 a_{1} \cdot a_{2} \xi_{3} \slsh{k_{1}} \xi_{4} 
	+  4 a_{1} \cdot a_{3} \xi_{2} \slsh{k_{3}} \xi_{4} -4 a_{1} \cdot a_{2} \xi_{3} \slsh{k_{2}} \xi_{4} \nn\\
	+ & 24 a_{2} \cdot a_{3} \xi_{1} \slsh{k_{3}} \xi_{4} -24 a_{2} \cdot a_{3} \xi_{1} \slsh{k_{2}} \xi_{4} 
	+  2 \xi_{2} [\slsh{k_{1}},\slsh{a_{1}}] \slsh{a_{3}} \xi_{4} -  2 \xi_{3} [\slsh{k_{1}},\slsh{a_{1}}] \slsh{a_{2}} 		\xi_{4} \nn\\
	+& (k_{2}^{a} a_{2}^{b} a_{3}^{c} - k_{3}^{a} a_{3}^{b} a_{2}^{c} ) \xi_{1} \g \xi_{4} 
	+    (16 k_{1}^{a} a_{1}^{b} a_{3}^{c} + 4 k_{3}^{a} a_{3}^{b} a_{1}^{c} ) \xi_{2} \g \xi_{4} 
	-  (16 k_{1}^{a} a_{1}^{b} a_{2}^{c} \nn\\
+ &4 k_{2}^{a} a_{2}^{b} a_{1}^{c} ) \xi_{3} \g \xi_{4} \Big\} \Big] 
 - (2,1,3) - (3,2,1) 
\end{align}

also equal to 

\beal 
\frac{i\al'}{5760} \Big[ \frac{\Pi(z_{ij})}{z_{1}-z_{4}} & \Big\{ 
	2\xi_{1} \slsh{a_{2}} \xi_{4} \, k_{2} \cdot a_{3}  \, 
	+  2\xi_{2} \slsh{a_{3}} \xi_{4} \, k_{4} \cdot a_{1}  \,
	+  a_{2} \cdot a_{3} \xi_{1} \slsh{k_{3}} \xi_{4}   
	+ \xi_{3} \slsh{a_{2}} \slsh{a_{1}} \slsh{k_{1}} \xi_{4} \Big\} \Big] \nn \\
+{\rm cyclic}
\end{align}

The contribution from $\beta_{4},\g_{4},\delta_{4}$ is

\beal 
\frac{i\al'}{5760} \Big[ \frac{\Pi(z_{ij})}{z_{3}-z_{4}} & \Big\{ 
	\xi_{1} \slsh{a_{3}} \xi_{2} \, k_{3} \cdot a_{4}  \, 
-\xi_{1} \slsh{a_{4}} \xi_{2} \, k_{4} \cdot a_{3}  \, 
	+  2\xi_{2} \slsh{a_{1}} \xi_{3} \, k_{3} \cdot a_{4}  \,
	+  a_{3} \cdot a_{4} \xi_{1} \slsh{k_{4}} \xi_{2}   
	+ \xi_{1} \slsh{a_{2}} \slsh{a_{4}} \slsh{k_{4}} \xi_{3} \Big\} \Big] \nn \\
+{\rm cyclic}
\end{align}

\section{Integrals}\label{integrals}

All integrals are reduced to hypergeometric functions using the formula 
\begin{align}
\int_{0}^{1} \, dt \, t^{b-1} (1-t)^{c-b-1} (1-tz)^{-a} & = \frac{\Gamma(b) \Gamma(c-b)}{\Gamma(c)} \, _{2}F_{1} (a,b;c;z) \,, \nn
\end{align} 
which is a single-valued analytic function of $z$ with a cut on the real axis from 1 to $\infty$. We use the notations 
\begin{align} 
%	 a = 2 k_{14}=2 k_{23}=-t \,; b = 2 k_{13}= 2 k_{24}=-u \,;& \, c= 2 k_{12}= 2 k_{34}=-s \,; \nn\\
	  y =  & \frac{z_{2}-z_{1}}{z_{3}-z_{1}}\nn \\ 
	 %z_{2}-z_{1}=x_{1} \,; z_{3}-z_{1} = x_2 \,. 		\nn\\ 
	%& x = z_{4}-z_{1} \,; 
	%\text{on-shell}  \quad a+b+c = 0&  \nn\\
	 p(y) =|z_{2}-z_{1}|^{ c} \, | z_{3}-z_{1}|^{ b} \, | z_{3}-z_{2}|^{ a} & = \frac{y^{c}}{(1-y)^{b+c}} ~.\nn
\end{align} 
The positions of the vertices are ordered so that $z_{1}<z_{2} <z_{3}$, hence $0<y<1$. The formulae below are not valid outside this range. 
The integrals needed for the amplitudes are the following:

\begin{align} 
	I_{1}& (a,b,c;z_{r}) \equiv p(y) \int_{-\infty}^{\infty} \frac{dz_{4}}{z_{1}-z_{4}} \, |z_{1}-z_{4}|^{a} \, |z_{2}-			z_{4}|^b \, |z_{3}-z_{4}|^c = \nn\\
	= & \,   \left( c + a  \, _{2}F_{1}(1,b; 1-c; y) \right)  \GV(a,b,c)
%		\left\{   \frac{\Gamma(b) \Gamma(-b-c)}{\Gamma(-c)} 
%			+\frac{c}{b} \, \frac{\Gamma(-b-c) \Gamma(c)}{\Gamma(-b)} 
%			- \frac{c}{b+c} \frac{\Gamma(b) \Gamma(c)}{\Gamma(b+c)} 
%		\right\} 
\nn \\
	&+ p(y) ( - \Gamma(-c) \Gamma(1+c) - \psi(-c) + \psi(1+c)) \, ; \nn	\\
	I_{2}& (a,b,c;z_{r}) \equiv p(y) \int_{-\infty}^{\infty} \frac{dz_{4}}{z_{2}-z_{4}} \, |z_{1}-z_{4}|^{a} \, |z_{2}-z_{4}|^b 		\, |z_{3}-z_{4}|^c = \nn\\
	= & \,   a \, _{2}F_{1}(1,b; 1-c; y)  \GV(a,b,c)
%		\left\{ - \frac{b+c}{c} \,  \frac{\Gamma(b) \Gamma(-b-c)}{\Gamma(-c)}  
%		- \frac{b+c}{b} \,  \frac{\Gamma(-b-c) \Gamma(c)}{\Gamma(-b)} 
%		+ \frac{\Gamma(b) \Gamma(c)}{\Gamma(b+c)} 
%		\right\} 
\nn \\
	&+p(y) ( - \Gamma(-c) \Gamma(1+c) - \psi(-c) + \psi(1+c)) \, ; \nn \\
	I_{3}& (a,b,c;z_{r}) \equiv p(y) \int_{-\infty}^{\infty} \frac{dz_{4}}{z_{3}-z_{4}} \, |z_{1}-z_{4}|^{a} \, |z_{2}-z_{4}|^b 		\, |z_{3}-z_{4}|^c = \nn\\
	= & \,   (a \, _{2}F_{1}(1,b; 1-c; y) -a)   \GV(a,b,c)
%	\left\{ - \frac{b+c}{c} \,  \frac{\Gamma(b) \Gamma(-b-c)}{\Gamma(-c)}  
%		- \frac{b+c}{b} \,  \frac{\Gamma(-b-c) \Gamma(c)}{\Gamma(-b)} 
%		+ \frac{\Gamma(b) \Gamma(c)}{\Gamma(b+c)} 
%		\right\} 
	\nn \\
	&+p(y)( -\Gamma(-c) \Gamma(1+c) 	- \psi(-c) + \psi(1+c) )\, , 
%	
%	+ \psi(1-b-c) - \psi(b+c)  + \frac{\Gamma (1-c) \Gamma (-b-c+1)
%	 \Gamma (c) \Gamma(b+c)}{\Gamma (1-b) \Gamma (b)} \, . \nn
\end{align}

where 
\begin{align}
\GV(a,b,c) &:= \frac{\Gamma(a) \Gamma(b)}{\Gamma(1-c)} + \frac{\Gamma(a)\Gamma(c)}{\Gamma(1-b)} + \frac{\Gamma(b)\Gamma(c)}{\Gamma(1-a)} \nn\\
&:=  G(a,b) + G(a,c) + G(b,c) \,.
\end{align}
On shell one has $a+b+c = 0$, and 
\beal 
\GV(a,b,c)& \sim -\frac{\pi^{2}}{2} + O(\al^{\prime 2})~.
\end{align}
One can easily check that the combination $\sum_{i} \al_{i} I_{i}$ is $y$-independent and hence $SL(2,\mathbb{R})$-invariant iff $\sum \al_{i}=0$. In fact, 
\beal
I_{1}-I_{2} = c, ~ I_{1} - I_{3} = -b, ~ I_{2}-I_{3} = a \,.
%\sum_{i} \al_{i} I_{i} = \al_{1} \, c - \al_{3} \, a = \al_{2} a - \al_{1} b = \al_{3} b - \al_{2} c \,.
\end{align}

\vfill\break

%%%%%%%%%
\section{Traces}
\label{apptraces}

The $u$, $v$ coefficients in equations (\ref{111}),(\ref{thelagrangian}), are defined as follows:
\begin{alignat}{2}
u_{ij}{}^{mn  pqm'n' p'q'; a_1\dots a_M;  b_1\dots b_N}& := \nn\\
-32\frac{c_Mc_N}{M!N!}\Big\{
 &2g^{mq}g^{m'q'}\delta_{i}^{p}\delta_{j}^{(n'}\delta_{k}^{p')}
<\g^n\g^{a_1\dots a_M}\g^k\g^{b_1\dots b_N}>
\varepsilon_N\nn\\
-&g^{m'q'}\delta_{i}^{q}\delta_{j}^{(n'}\delta_{k}^{p')}
<\g^{mnp}\g^{a_1\dots a_M}\g^k\g^{b_1\dots b_N}>
(\varepsilon_M+\varepsilon_N)\nn\\
+&\frac{1}{2}\delta_i^{[q|}\delta_j^{q'}
<\g^{|mnp]}\g^{a_1\dots a_M}\g^{m'n'p'}\g^{b_1\dots b_N}>
\varepsilon_N
\Big\}
~
\end{alignat}
%
%\frac{3}{4}
%g^{m'q}\delta_{i}^{[n'}\delta_{j}^{p'}\delta_{k}^{q']}
%<\g^{mnp}\g^{a_1\dots a_M}\g^k\g^{b_1\dots b_N}>
%\varepsilon_N\nn\\
%-&\frac{1}{8}g_{ij}
%<\g^{mnp}\g^{a_1\dots a_M}\g^{m'n'p'}\g^{b_1\dots b_N}>
%\varepsilon_N
%
and
\begin{alignat}{2}
v^{a_1\dots a_M;  b_1\dots b_N;   c_1\dots c_P;   d_1\dots d_Q}& := \nn\\
\frac{32}{9}\frac{c_Mc_Nc_Pc_Q}{M!N!P!Q!}\Big\{
 &<\g^{a_1\dots a_M}\g_q\g^{b_1\dots b_N}\g_n\g^{c_1\dots c_P}\g^q\g^{d_1\dots d_Q}\g^n>
\varepsilon_N\varepsilon_Q\nn\\
-&<\g^{a_1\dots a_M}\g_q\g^{b_1\dots b_N}\g_n><\g^{c_1\dots c_P}\g^q\g^{d_1\dots d_Q}\g^n>
\varepsilon_N\varepsilon_Q\nn\\
-&5<\g^{a_1\dots a_M}\g_q\g^{c_1\dots c_P}\g_n\g^{b_1\dots b_N}\g^q\g^{d_1\dots d_Q}\g^n>
\varepsilon_P\varepsilon_Q\nn\\
+&4<\g^{a_1\dots a_M}\g_q\g^{c_1\dots c_P}\g_n><\g^{b_1\dots b_N}\g^q\g^{d_1\dots d_Q}\g^n>
\varepsilon_P\varepsilon_Q\nn\\
+&<\g^{a_1\dots a_M}\g_q\g^{c_1\dots c_P}\g_n\g^{d_1\dots d_Q}\g^q\g^{b_1\dots b_N}\g^n>
\varepsilon_P\varepsilon_N
\Big\}
~,
\end{alignat}
%
%and
%%
%\begin{alignat}{2}
%w_{ij}{}^{rsmn  pq; a_1\dots a_M;  b_1\dots b_N} := 
%\frac{64}{3}\frac{c_Mc_N\varepsilon_N}{M!N!}\Big\{
% 2&g^{mq}\delta_{i}^{s}\delta_{j}^{(n|}
%<\g^r\g^{a_1\dots a_M}\g^{|p)}\g^{b_1\dots b_N}>\nn\\
%+&\delta_{i}^{s}\delta_{j}^{[m}
%<\g^r\g^{a_1\dots a_M}\g^{|npq]}\g^{b_1\dots b_N}>
%\Big\}
%~,
%\end{alignat}
%%
where 
\beal
\varepsilon_M:=(-1)^{\frac{1}{2}M(M-1)}~,
\end{align}
and $c_M$ was defined in (\ref{fexp}). 
The explicit form of the traces above can readily obtained using a symbolic program 
for the manipulation of $\g\acute{\alpha}\mu\mu\al$ - matrices, e.g. \cite{gran}\footnote{
In using \cite{gran}, care should be taken to include the contribution of 
the totally-antisymmetric 
epsilon tensor in ten dimensions, which is not automatically taken care of by the program.}. 
Note that the result thus obtained will not necessarily be 
expressed in a basis of independent invariants; additional 
manipulations are needed if one wishes to bring the result to 
a form involving a minimal number of terms.

\section{$(\partial F_{(1)})^2R^2$}
\label{last}

To illustrate the procedure, let us examine the $(\partial F_{(1)})^2R^2$ 
couplings in more detail. 
First note that in the linearized approximation the equation of motion for $F_{(1)}$ reads 
$\partial^mF_m=0$. In addition, $F_{(1)}$ must be closed by the Bianchi identities. These two 
conditions are equivalent 
to the statement that $\partial_mF_n$ is a traceless symmetric tensor. In the 
Dynkin notation for $D_5$:
$$
\partial_m F_n\sim (20000)~.
$$
Similarly, at the linearized level, the equation of motion for 
the graviton reads $R_{mn}=0$. In addition, the Riemann tensor obeys the 
Bianchi identities $R_{[mnp]q}=0$. Together with the symmetry of the Riemann tensor 
$R_{mnpq}=R_{pqmn}$, these constraints can be expressed compactly as
$$
R_{mnpq}\sim (02000)~.
$$
It follows that in the case 
at hand there are exactly five inequivalent scalars which can be constructed.
In Dynkin notation:
$$
(\partial F_{(1)})^2R^2\sim  (20000)^{2\otimes_s}\otimes (02000)^{2\otimes_s}
\sim 5\times(00000)\oplus\dots
~.
$$
Explicitly, we can choose a basis $I_1,\dots I_5$ of these five scalars as follows
\beal
I_1&:=\partial_{m}F^n\partial_pF^q R^{imjp}R_{injq} \nn\\
I_2&:=\partial_{m}F_n\partial^pF^q R^{imjn}R_{ipjq}\nn\\
I_3&:=\partial_{m}F^n\partial_pF^q R^{mpij}R_{ijnq}\nn\\
I_4&:=\partial_{m}F_i\partial^iF^n R^{mjkl}R_{njkl}\nn\\
I_5&:=\partial_{i}F_j\partial^iF^j R^{klmn}R_{klmn}~.
\end{align}
In the linearized approximation around flat space we have in addition: 
$R_{mn}{}^{pq}\sim \partial_{[m}\partial^{[p}h_{n]}{}^{q]}$. Taking this 
into account, it is straightforward to show that in this approximation 
the invariants above are not independent, but obey 
\beal
I_1-I_2+\frac{1}{2}I_3+I_4-\frac{1}{8}I_5=0~.
\label{lindep}
\end{align}
As we have argued in section \ref{53}, in the linearized approximation around flat space 
there is a relation
\beal
{R}_{mnm'n'}{R}_{pqp'q'}
<\g^{[mnp}\partial^{q]}F\g^{m'n'p'}\partial^{q'}F^{Tr}>
={R}_{mnm'n'}{R}_{pqp'q'}<F\g^{[m'n'p'}\partial^{q']}\partial^{q}F^{Tr}\g^{mnp}>
~.
\end{align}
This can be explicitly verified in the case at hand: a straightforward computation yields
\beal
{R}_{mnm'n'}{R}_{pqp'q'}
<\g^{[mnp}\partial^{q]}F\g^{m'n'p'}\partial^{q'}F^{Tr}>&=64(I_1-I_2+\frac{1}{2}I_3+\frac{1}{2}I_4)\nn\\
{R}_{mnm'n'}{R}_{pqp'q'}
<F\g^{[m'n'p'}\partial^{q']}\partial^{q}F^{Tr}\g^{mnp}>&=
-64(I_1-I_2+\frac{1}{2}I_3+\frac{3}{2}I_4-\frac{1}{4}I_5)~.
\end{align}
The expressions on the right-hand sides of the two equations above can indeed be seen to 
be equal, when (\ref{lindep}) is taken into account. 

The couplings $(\partial F_{(1)})^2R^2$ are related to the $(\partial D)^2R^2$ 
couplings, coming from $t_8t_8\widehat{R}^4$, by $SL(2,\mathbb{Z})$ duality. We have 
directly verified 
that the sum of the two contributions is indeed $SL(2,\mathbb{Z})$ invariant, as expected.

%%%%%%%%%%%%%%%%%%%%%%%%%%%%%%%%%%%%%%%%%%%%
%
% Bibliography
%
%%%%%%%%%%%%%%%%%%%%%%%%%%%%%%%%%%%%%%%%%%%%


\begin{thebibliography}{99}

\bibitem{ba}
N.~Berkovits,
``Super-Poincare covariant quantization of the superstring,'' 
JHEP {\bf 0004} (2000) 018, hep-th/0001035.


\bibitem{bb}
  N.~Berkovits,
  ``ICTP lectures on covariant quantization of the superstring,''
  hep-th/0209059.


%
\bibitem{sym}
  B.~E.~W.~Nilsson,
 ``Off-shell fields for the 10-dimensional supersymmetric Yang-Mills theory'', 
G\"{o}teborg-ITP-81-6;  
``Pure spinors as auxiliary fields in the ten-dimensional supersymmetric
  Yang-Mills theory,''
  Class.\ Quant.\ Grav.\  {\bf 3} (1986) L41.



\bibitem{purehoweten} P.~S.~Howe, ``Pure spinors lines in superspace and ten-dimensional supersymmetric theories,'' Phys.\ Lett.\ B {\bf 258}, 141 (1991) [Addendum-ibid.\ B {\bf 259}, 511 (1991)]. 
\bibitem{purehoweeleven} P.~S.~Howe, ``Pure spinors, function superspaces and supergravity theories in ten-dimensions and eleven-dimensions,'' Phys.\ Lett.\ B {\bf 273} (1991) 90. 



\bibitem{b}
N.~Berkovits and B.~C.~Vallilo,
 ``Consistency of super-Poincare covariant superstring tree amplitudes,''
  JHEP {\bf 0007} (2000) 015, hep-th/0004171.




\bibitem{bc}
  N.~Berkovits,
  ``Relating the RNS and pure spinor formalisms for the superstring,''
  JHEP {\bf 0108} (2001) 026, hep-th/0104247.


\bibitem{bca}
  N.~Berkovits and O.~Chandia,
  ``Lorentz invariance of the pure spinor BRST cohomology for the
  superstring,''
  Phys.\ Lett.\ B {\bf 514} (2001) 394, hep-th/0105149.



\bibitem{bd}
N.~Berkovits,
``Covariant multiloop superstring amplitudes,'' hep-th/0410079.




\bibitem{bda}
N.~Berkovits,
 ``Multiloop amplitudes and vanishing theorems using the pure spinor
  formalism for the superstring,''
  JHEP {\bf 0409} (2004) 047, hep-th/0406055.


\bibitem{be}
  N.~Berkovits and D.~Z.~Marchioro,
  ``Relating the Green-Schwarz and pure spinor formalisms for the
  superstring,''
  JHEP {\bf 0501} (2005) 018, hep-th/0412198.


\bibitem{bf}
  N.~Berkovits and C.~R.~Mafra,
  ``Equivalence of two-loop superstring amplitudes in the pure spinor and RNS
  formalisms,''
 Phys.\ Rev.\ Lett.\  {\bf 96} (2006) 011602, hep-th/0509234.


%\cite{Aisaka:2005vn}
\bibitem{Aisaka:2005vn}
  Y.~Aisaka and Y.~Kazama,
  ``Origin of pure spinor superstring,''
  JHEP {\bf 0505} (2005) 046, hep-th/0502208.
  


\bibitem{giusa}
P.~A.~Grassi, G.~Policastro, M.~Porrati and P.~Van Nieuwenhuizen,
``Covariant quantization of superstrings without pure spinor constraints,''
JHEP {\bf 0210} (2002) 054, hep-th/0112162.


\bibitem{giusb}
  P.~A.~Grassi, G.~Policastro and P.~van Nieuwenhuizen,
  ``The covariant quantum superstring and superparticle from their classical
  actions,''
  Phys.\ Lett.\ B {\bf 553} (2003) 96, hep-th/0209026.

  %\cite{Grassi:2003kq}
\bibitem{Grassi:2003kq}
  P.~A.~Grassi, G.~Policastro and P.~van Nieuwenhuizen,
  ``The quantum superstring as a WZNW model,''
  Nucl.\ Phys.\ B {\bf 676} (2004) 43, hep-th/0307056.

%\cite{Grassi:2002tz}
\bibitem{Grassi:2002tz}
  P.~A.~Grassi, G.~Policastro and P.~van Nieuwenhuizen,
  ``The massless spectrum of covariant superstrings,''
  JHEP {\bf 0211} (2002) 001, hep-th/0202123.


%\cite{Berkovits:2005bt}
\bibitem{Berkovits:2005bt}
  N.~Berkovits,
  ``Pure spinor formalism as an N = 2 topological string,''
  JHEP {\bf 0510} (2005) 089, hep-th/0509120.
  

%\cite{Oda:2005sd}
\bibitem{Oda:2005sd}
  I.~Oda and M.~Tonin,
  ``Y-formalism in pure spinor quantization of superstrings,''
  Nucl.\ Phys.\ B {\bf 727} (2005) 176, hep-th/0505277.
  
\bibitem{Gaona:2005yw}
  A.~Gaona and J.~A.~Garcia,
  ``BFT embedding of the Green-Schwarz superstring and the pure spinor formalism,''
  JHEP {\bf 0509} (2005) 083
  [arXiv:hep-th/0507076].


%
\bibitem{grosssloan}
D.~J.~Gross, J.~H.~Sloan, ``The Quartic Effective action for the Heterotic String'', 
Nucl.\ Phys.\ B {\bf 291} (1987) 41.

\bibitem{kehagiasa} A.~Kehagias and H.~Partouche, ``The exact quartic effective action for the type IIB superstring'', Phys.\ Lett.\ B {\bf 422} (1998) 109. 
\bibitem{kehagiasb} A.~Kehagias and H.~Partouche, ``D-instanton corrections as (p,q)-string effects and non-renormalization theorems,'' Int.\ J.\ Mod.\ Phys.\ A {\bf 13} (1998) 5075. 




  \bibitem{klt}
H.~Kawai, D.~C.~Lewellen and S.~H.~H.~Tye,
``A Relation Between Tree Amplitudes Of Closed And Open Strings,''
Nucl.\ Phys.\ B {\bf 269} (1986) 1.

    
  %\cite{Bern:2002kj}
\bibitem{Bern:2002kj}
  Z.~Bern,
  ``Perturbative quantum gravity and its relation to gauge theory,''
  Living Rev.\ Rel.\  {\bf 5} (2002) 5, gr-qc/0206071.  
  
  \bibitem{gt}
P.~A.~Grassi and L.~Tamassia,
``Vertex operators for closed superstrings,'' 
JHEP {\bf 0407} (2004) 071, hep-th/0405072.

\bibitem{mac}
I.~N.~McArthur,
``Superspace Normal Coordinates,'' 
Class.\ Quant.\ Grav.\  {\bf 1} (1984) 233.


\bibitem{curved}
D.~Tsimpis,
``Curved 11D supergeometry,''
JHEP {\bf 0411} (2004) 087, hep-th/0407244.

%
\bibitem{chiral}
K.~Peeters, P.~Vanhove and A.~Westerberg,
``Chiral splitting and world-sheet gravitinos in higher-derivative string
amplitudes,''
Class.\ Quant.\ Grav.\  {\bf 19} (2002) 2699, hep-th/0112157.
%

\bibitem{pvwreview}
 K.~Peeters, P.~Vanhove and A.~Westerberg,
 ``Supersymmetric higher-derivative actions in ten and eleven dimensions,  the
 associated superalgebras and their formulation in superspace,''
  Class.\ Quant.\ Grav.\  {\bf 18} (2001) 843, hep-th/0010167.

%
\bibitem{hw}
P.~S.~Howe and P.~C.~West,
``The Complete N=2, D = 10 Supergravity,''
Nucl.\ Phys.\ B {\bf 238} (1984) 181.

%
\bibitem{nilsson}
 B.~E.~W.~Nilsson and A.~K.~Tollsten,
  ``Supersymmetrization of $\zeta (3) R_{\mu \nu \rho \sigma}^4$ in superstring
  theories,''
  Phys.\ Lett.\ B {\bf 181} (1986) 63.
%
\bibitem{gps}
  M.~B.~Green, K.~Peeters and C.~Stahn,
  ``Superfield integrals in high dimensions,''
  JHEP {\bf 0508} (2005) 093, hep-th/0506161.
%
\bibitem{pw}
K.~Peeters and A.~Westerberg,
``The Ramond-Ramond sector of string theory beyond leading order,''
Class.\ Quant.\ Grav.\  {\bf 21} (2004) 1643, hep-th/0307298.

%
\bibitem{aticksen}
J.~J.~Atick and A.~Sen,
``Covariant One Loop Fermion Emission Amplitudes In Closed String Theories,''
  Nucl.\ Phys.\ B {\bf 293} (1987) 317.



%
\bibitem{giani}
F.~Giani and M.~Pernici,
``N=2 Supergravity In Ten-Dimensions,''
  Phys.\ Rev.\ D {\bf 30} (1984) 325.


\bibitem{bh}
N.~Berkovits and P.~S.~Howe, unpublished.




%
\bibitem{skenderis}
S.~de Haro, A.~Sinkovics and K.~Skenderis,
``On a supersymmetric completion of the $R^4$ term in IIB supergravity,''
Phys.\ Rev.\ D {\bf 67} (2003) 084010, hep-th/0210080.




%


\bibitem{cgnt} 
M.~Cederwall, U.~Gran, B.~E.~W.~Nilsson and D.~Tsimpis, 
  ``Supersymmetric corrections to eleven-dimensional supergravity,''
  JHEP {\bf 0505}, 052 (2005), hep-th/0409107.

%
\bibitem{lthree}
  D.~Tsimpis,
  ``11D supergravity at ${\cal O}(l^3)$,''
  JHEP {\bf 0410}, 046 (2004), hep-th/0407271.

%
\bibitem{witten}
 E.~Witten,
 ``On flux quantization in M-theory and the effective action,''
  J.\ Geom.\ Phys.\  {\bf 22} (1997) 1, hep-th/9609122.

%

\bibitem{roufatokavli}
  Y.~Hyakutake and S.~Ogushi,
  ``Higher derivative corrections to eleven dimensional supergravity via local
  supersymmetry,''
  JHEP {\bf 0602} (2006) 068, hep-th/0601092.

%

\bibitem{cgnn}
M.~Cederwall, U.~Gran, M.~Nielsen and B.~E.~W.~Nilsson,
``Manifestly supersymmetric M-theory,'' 
JHEP {\bf 0010} (2000) 041, hep-th/0007035; 
  ``Generalised 11-dimensional supergravity,'' hep-th/0010042.

\bibitem{ht}
P.~S.~Howe and D.~Tsimpis,
``On higher-order corrections in M theory,''
JHEP {\bf 0309} (2003) 038, hep-th/0305129.

%
\bibitem{md}
M.~J.~Duff, J.~T.~Liu and R.~Minasian,
``Eleven-dimensional origin of string / string duality: A one-loop test,''
Nucl.\ Phys.\ B {\bf 452} (1995) 261, hep-th/9506126.
%
\bibitem{deroo}
M.~de Roo, M.~G.~C.~Eenink, 
``The effective action for the 4-point functions in abelian open superstring  theory'', hep-th/0307211.


  
\bibitem{gran}
U.~Gran,
``GAMMA: A Mathematica package for performing Gamma-matrix algebra and  Fierz
transformations in arbitrary dimensions,'' hep-th/0105086.



\bibitem{gva} L.~Anguelova, P.~A.~Grassi and P.~Vanhove, ``Covariant one-loop amplitudes in D = 11,'' Nucl.\ Phys.\ B {\bf 702} (2004) 269.
\bibitem{gvb} P.~A.~Grassi and P.~Vanhove,  ``Topological M theory from pure spinor formalism,'' hep-th/0411167. 
\bibitem{mafra} C.~R.~Mafra, ``Four-point one-loop amplitude computation in the pure spinor formalism,'' JHEP {\bf 0601} (2006) 075. 




\bibitem{cnt}
M.~Cederwall, B.~E.~W.~Nilsson and D.~Tsimpis,
``Spinorial cohomology and maximally supersymmetric theories,''
  JHEP {\bf 0202}, 009 (2002), hep-th/0110069; 
  ``The structure of maximally supersymmetric Yang-Mills theory:  Constraining
  higher-order corrections,''
  JHEP {\bf 0106}, 034 (2001), hep-th/0102009.



\bibitem{gg}
M.~B.~Green and M.~Gutperle,
``Effects of D-instantons,''
Nucl.\ Phys.\ B {\bf 498} (1997) 195, hep-th/9701093.



\end{thebibliography}
\end{document}